\documentclass[10pt,aps,pra,onecolumn]{revtex4-2}
\usepackage{amsmath}
\usepackage{amssymb}
\usepackage{graphicx}
\usepackage{hyperref}
\usepackage[usenames,dvipsnames]{color}
\usepackage[portuguese]{babel}
\usepackage[T1]{fontenc}

\begin{document}

\title{Resson\^{a}ncias de um e m\'{u}ltiplos f\'{o}tons na intera\c{c}\~{a}%
o entre luz e \'atomos \\
(One- and multiphoton resonances in the light-atom interaction)}
\author{Alexandre P. Costa}
\thanks{ORCID: 0000-0002-5456-7251}
\author{Alexandre Dodonov}
\thanks{ORCID: 0000-0002-7142-7453}
\email{Email: adodonov@unb.br}
\affiliation{Centro Internacional de Física, Instituto de F\'{\i}sica, Universidade de Bras\'{\i}lia, Caixa Postal 04455, CEP
70919-970 Bras\'{\i}lia, DF, Brasil}
\begin{abstract}
A intera\c{c}\~{a}o entre sistemas at\^{o}micos e campos eletromagn\'{e}%
ticos \'{e} central na f\'{\i}sica moderna e nas tecnologias qu\^{a}nticas
emergentes. Os modelos de Rabi, em suas vers\~{o}es semicl\'{a}ssica e qu%
\^{a}ntica, fornecem a descri\c{c}\~{a}o mais simples e fundamental dessa
intera\c{c}\~{a}o. Neste trabalho, apresentamos uma dedu\c{c}\~{a}o resumida
de ambos os modelos e mostramos como surgem resson\^{a}ncias de um e m\'{u}%
ltiplos f\'{o}tons no regime semicl\'{a}ssico. Em seguida, analisamos como
essas resson\^{a}ncias se manifestam no modelo de Rabi qu\^{a}ntico,
discutindo semelhan\c{c}as e diferen\c{c}as em rela\c{c}\~{a}o \`{a} descri%
\c{c}\~{a}o cl\'{a}ssica. Damos aten\c{c}\~{a}o especial \`{a} resson\^{a}%
ncia de tr\^{e}s f\'{o}tons, fen\^{o}meno normalmente negligenciado em
livros-texto por sua fraqueza relativa, mas que \'{e} intr\'{\i}nseco \`{a}
intera\c{c}\~{a}o radia\c{c}\~{a}o--mat\'{e}ria. Nosso objetivo \'{e}
oferecer uma refer\^{e}ncia did\'{a}tica acess\'{\i}vel para estudantes e
pesquisadores interessados em \'{O}ptica Qu\^{a}ntica e Informa\c{c}\~{a}o Qu%
\^{a}ntica, com \^{e}nfase nos fundamentos dos modelos de Rabi.\\

(The interaction between atomic systems and electromagnetic fields is central to modern physics and emerging quantum technologies. The Rabi models, in their semiclassical and quantum versions, provide the simplest and most fundamental description of this interaction. In this work, we present a concise derivation of both models and show how one- and multiphoton resonances arise in the semiclassical regime. We then analyze how these resonances manifest in the quantum Rabi model, discussing similarities and differences in relation to the classical description. Special attention is given to the three-photon resonance, a phenomenon usually neglected in textbooks due to its relative weakness, but which is intrinsic to the radiation–matter interaction. Our goal is to offer an accessible didactic reference for students and researchers interested in Quantum Optics and Quantum Information, with an emphasis on the fundamentals of the Rabi models.)
\end{abstract}

\maketitle

\noindent\textbf{Palavras-chave:} Modelo de Rabi; Óptica Quântica; Informação Quântica; interação radiação-matéria; ressonância multifotônica.


\section{Introdu\c{c}\~{a}o}

Desde o in\'{\i}cio do s\'{e}culo XX, a intera\c{c}\~{a}o entre a radia\c{c}%
\~{a}o eletromagn\'{e}tica e sistemas at\^{o}micos tem sido um dos temas
centrais na f\'{\i}sica moderna. A compreens\~{a}o de como a luz interage
com a mat\'{e}ria est\'{a} na base de diversas teorias fundamentais, como
Eletrodin\^{a}mica Qu\^{a}ntica, \'{O}ptica N\~{a}o-linear e \'{O}ptica Qu%
\^{a}ntica, e possibilita a descri\c{c}\~{a}o precisa de fen\^{o}menos como
cria\c{c}\~{a}o e destrui\c{c}\~{a}o de f\'{o}tons, transi\c{c}\~{o}es at%
\^{o}micas, emiss\~{o}es espont\^{a}nea e estimulada, fluoresc\^{e}ncia,
transpar\^{e}ncia induzida eletromagneticamente, fotodetec\c{c}\~{a}o,
regimes fortes de intera\c{c}\~{a}o radia\c{c}\~{a}o--mat\'{e}ria, intera%
\c{c}\~{a}o entre n\'{\i}veis eletr\^{o}nicos de \'{\i}ons e f\^{o}nons
vibracionais, etc \cite{shore,boyd,scully,ultra1,ultra2,sol14,sol15,sol16,cohen}.
Nesse contexto, os modelos de Rabi, semicl\'{a}ssico e qu\^{a}ntico,
desempenham um papel fundamental na descri\c{c}\~{a}o da intera\c{c}\~{a}o
entre \'{a}tomos e o campo eletromagn\'{e}tico, ao fornecer a descri\c{c}%
\~{a}o mais b\'{a}sica de um sistema qu\^{a}ntico interagindo com um campo
bos\^{o}nico \cite{merlin,rev}.

Introduzido originalmente por Isidor Rabi nos anos 1930 \cite{rabi1,rabi2},
o modelo de Rabi semicl\'{a}ssico descreve a din\^{a}mica de um sistema qu%
\^{a}ntico de dois n\'{\i}veis sob a influ\^{e}ncia de um campo cl\'{a}ssico
oscilante \cite{bloch}, e \'{e} provavelmente um dos primeiros modelos que
se aprende ao estudar a intera\c{c}\~{a}o entre luz e mat\'{e}ria nos cursos
de \'{O}ptica N\~{a}o-linear ou F\'{\i}sica At\^{o}mica~\cite%
{boyd,scully,shore}. O trabalho de Rabi foi motivado pela necessidade de
descrever o comportamento de spins nucleares em campos magn\'{e}ticos
alternados, mas com o passar do tempo, o modelo passou a ser uma ferramenta
te\'{o}rica fundamental em sistemas t\~{a}o diversos como resson\^{a}ncia
magn\'{e}tica nuclear, \'{a}tomos frios, estados de impurezas em isolantes,
pontos qu\^{a}nticos, circuitos supercondutores, \'{a}tomos de Rydberg,
polaritons em cavidades, sistemas optomecânicos, etc. \cite{merlin,rev,wang-PRA-2024}, e diversas solu\c{c}\~{o}es
anal\'{\i}ticas foram encontradas em diferentes regimes de par\^{a}metros 
\cite{graham,munz,Liu,Lu,ashhab,castanos}.

Com o avan\c{c}o da F\'{\i}sica Qu\^{a}ntica, especialmente a partir da d%
\'{e}cada de 1960, tornou-se claro que muitos fen\^{o}menos n\~{a}o podiam
ser completamente compreendidos sem levar em conta a quantiza\c{c}\~{a}o do
pr\'{o}prio campo eletromagn\'{e}tico. Isso levou ao desenvolvimento do
modelo de Rabi qu\^{a}ntico, no qual tanto o sistema de dois n\'{\i}veis
quanto o campo s\~{a}o tratados quanticamente \cite{rev,solano,braak}. Essa
abordagem est\'{a} na base da \'{O}ptica Qu\^{a}ntica e da Informa\c{c}\~{a}%
o Qu\^{a}ntica, e modelos derivados, como o modelo de Jaynes--Cummings, de
Tavis-Cummings ou de Dicke \cite{larson}, tornaram-se paradigmas para
estudar a intera\c{c}\~{a}o entre \'{a}tomos e f\'{o}tons individuais em
cavidades \'{o}pticas ou de microondas, circuitos supercondutores e
dispositivos de computa\c{c}\~{a}o qu\^{a}ntica \cite{mi,mat,calib,NMR,dai}.
Al\'{e}m do seu valor conceitual, os modelos de Rabi s\~{a}o ferramentas pr%
\'{a}ticas em experimentos de controle qu\^{a}ntico, manipula\c{c}\~{a}o de
estados at\^{o}micos, gera\c{c}\~{a}o controlada de f\'{o}tons, prepara\c{c}%
\~{a}o e detec\c{c}\~{a}o de estados qu\^{a}nticos, cria\c{c}\~{a}o e
manipula\c{c}\~{a}o de estados emaranhados, implementa\c{c}\~{a}o de portas l%
\'{o}gicas qu\^{a}nticas, etc \cite{larson}. Dado seu amplo alcance e
aplicabilidade, o estudo detalhado desses modelos --- tanto em sua vers\~{a}%
o semicl\'{a}ssica quanto plenamente qu\^{a}ntica --- \'{e} essencial para
qualquer programa de forma\c{c}\~{a}o em \'{O}ptica Qu\^{a}ntica, \'{O}ptica
N\~{a}o-linear e tecnologias qu\^{a}nticas \cite%
{nielsen,haroche,scully,knight,schleich,boyd,nifa}.

Neste artigo, vamos fazer uma deriva\c{c}\~{a}o resumida dos modelos de Rabi
semicl\'{a}ssico e qu\^{a}ntico, mostrar analiticamente como as resson\^{a}%
ncias de um e m\'{u}ltiplos f\'{o}tons surgem naturalmente no modelo semicl%
\'{a}ssico, e descrever a dinâmica do sistema nestas circunstâncias. Em seguida, vamos mostrar como essas resson\^{a}ncias aparecem
no modelo qu\^{a}ntico, discutindo as semelhan\c{c}as e diferen\c{c}as entre
as previs\~{o}es semicl\'{a}ssica e qu\^{a}ntica. Como as resson\^{a}ncias
multifot\^{o}nicas s\~{a}o fen\^{o}menos muito fracos, comparadas \`{a}
resson\^{a}ncia de um f\'{o}ton, elas n\~{a}o costumam ser abordadas nos
livros-texto de \'{O}ptica Qu\^{a}ntica e Informa\c{c}\~{a}o Qu\^{a}ntica.
Portanto, este trabalho tamb\'{e}m poder\'{a} ser usado como uma refer\^{e}%
ncia did\'{a}tica para explicar de forma simples este fen\^{o}meno inerente
\`a intera\c{c}\~{a}o radia\c{c}\~{a}o--mat\'{e}ria.

Este artigo est\'{a} estruturado da seguinte forma. Na se\c{c}\~{a}o \ref%
{deducao} deduzimos, de forma resumida, os modelos de Rabi a partir de
primeiros princ\'{\i}pios. Na se\c{c}\~{a}o \ref{SRM} encontramos as solu%
\c{c}\~{o}es aproximadas do modelo de Rabi semicl\'{a}ssico, e mostramos o
comportamento t\'{\i}pico do \'{a}tomo perto das resson\^{a}ncias. Na se\c{c}%
\~{a}o \ref{QRMz} descrevemos o modelo de Rabi qu\^{a}ntico e a sua rela\c{c}%
\~{a}o com o modelo semicl\'{a}ssico, e discutimos as semelhan\c{c}as e
diferen\c{c}as entre as previs\~{o}es semicl\'{a}ssicas e qu\^{a}nticas por
meio de exemplos concretos. Finalmente, na se\c{c}\~{a}o \ref{conc}
apresentamos as nossas conclus\~{o}es.

\section{Dedu\c{c}\~{a}o resumida dos modelos de Rabi\label{deducao}}

Os modelos de Rabi, semicl\'{a}ssico e qu\^{a}ntico, s\~{a}o abordados em in%
\'{u}meros livros de \'{O}ptica Qu\^{a}ntica e \'{O}ptica N\~{a}o-linear 
\cite{boyd,scully,schleich,knight,larson,haroche,orszag}. No entanto, grande
parte dos livros-texto n\~{a}o discute as resson\^{a}ncias multifot\^{o}%
nicas intr\'{\i}nsecas aos modelos de Rabi. Por isso, neste trabalho temos
como objetivo descrever um formalismo matem\'{a}tico bastante simples para
determinar as condi\c{c}\~{o}es para as resson\^{a}ncias multifot\^{o}nicas
e descrever a din\^{a}mica do sistema nestas circunst\^ancias.

Primeiro, vamos mostrar como os modelos de Rabi surgem naturalmente ao
considerar um el\'{e}tron ligado ao n\'{u}cleo at\^{o}mico na presen\c{c}a
do campo eletromagn\'{e}tico. Para simplificar, vamos considerar o \'{a}tomo
de Hidrog\^{e}nio, que consiste de um \'{u}nico el\'{e}tron (de carga $-e$)
ligado ao n\'{u}cleo de carga $+e$. Supondo que o n\'{u}cleo esteja parado,
na aus\^{e}ncia de campos externos, o Hamiltoniano do el\'{e}tron \'{e} dado
por%
\begin{equation}
\hat{H}_{0}=\frac{\mathbf{\hat{p}}^{2}}{2m}+V\left( \mathbf{\hat{r}}\right)
\,,  \label{h0}
\end{equation}%
onde 
\begin{equation}
V\left( \mathbf{\hat{r}}\right) =-\frac{1}{4\pi \varepsilon _{0}}\frac{e^{2}%
}{\left\vert \mathbf{\hat{r}}\right\vert }
\end{equation}%
\'{e} a energia potencial Coulombiana ($\varepsilon _{0}$ \'{e} a
permissividade el\'{e}trica do v\'{a}cuo), $\mathbf{\hat{r}}$ \'{e} o
operador-posi\c{c}\~{a}o do el\'{e}tron e $\mathbf{\hat{p}}$ \'{e} o
operador momentum linear. Na representa\c{c}\~{a}o de posi\c{c}\~{a}o, temos 
$\mathbf{\hat{p}=}-i\hbar \nabla $ (onde $\hbar $ \'{e} a constante reduzida
de Planck), e a fun\c{c}\~{a}o de onda do el\'{e}tron, $\psi \left( \mathbf{r%
},t\right) =\langle \mathbf{r}|\psi \left( t\right) \rangle $, obedece \`{a}
Equa\c{c}\~{a}o de Schr\"{o}dinger%
\begin{equation}
i\hbar \frac{\partial \psi \left( \mathbf{r},t\right) }{\partial t}=\hat{H}%
_{0}\psi \left( \mathbf{r},t\right) \,.
\end{equation}%
Os estados estacion\'{a}rios do el\'{e}tron s\~{a}o dados por%
\begin{equation}
\psi _{n}\left( \mathbf{r},t\right) =e^{-itE_{n}/\hbar }\psi _{n}\left( 
\mathbf{r}\right) \,,
\end{equation}%
onde $E_{n}$ e $\psi _{n}\left( \mathbf{r}\right) $ s\~{a}o as autoenergias
e as autofun\c{c}\~{o}es do Hamiltoniano $\hat{H}_{0}$, respectivamente,
obtidas da equa\c{c}\~{a}o de Schr\"{o}dinger independente do tempo%
\begin{equation}
\left( -\frac{\hbar ^{2}}{2m}\nabla ^{2}+V\left( \mathbf{r}\right) \right)
\psi _{n}\left( \mathbf{r}\right) =E_{n}\psi _{n}\left( \mathbf{r}\right) \,.
\end{equation}

Na presen\c{c}a do campo eletromagn\'{e}tico, na representa\c{c}\~{a}o de
posi\c{c}\~{a}o, o Hamiltoniano \'{e} dado por \cite{knight,schleich}%
\begin{equation}
\hat{H}=\frac{\left[ \mathbf{\hat{p}}+e\mathbf{A}\left( \mathbf{r},t\right) %
\right] ^{2}}{2m}-e\Phi \left( \mathbf{r},t\right) +V\left( \mathbf{r}%
\right)\, ,  \label{b4}
\end{equation}%
onde $\mathbf{A}\left( \mathbf{r},t\right) $ \'{e} o potencial vetorial e $%
\Phi \left( \mathbf{r},t\right) $ \'{e} o potencial escalar do campo
eletromagn\'{e}tico (na posi\c{c}\~{a}o $\mathbf{r}$ e no tempo $t$). O
campo el\'{e}trico e o campo magn\'{e}tico (indu\c{c}\~{a}o magn\'{e}tica,
para sermos mais rigorosos) s\~{a}o dados por%
\begin{equation}
\mathbf{E}\left( \mathbf{r},t\right) =-\nabla \Phi \left( \mathbf{r}%
,t\right) -\frac{\partial \mathbf{A}\left( \mathbf{r},t\right) }{\partial t}
\label{eq:field-ele}
\end{equation}%
\begin{equation}
\mathbf{B}\left( \mathbf{r},t\right) =\nabla \times \mathbf{A}\left( \mathbf{%
r},t\right) \,.
\label{eq:field-mag}
\end{equation}%
Note que estes campos s\~{a}o invariantes por transforma\c{c}\~{o}es de calibre (ver Ref. \cite{griffiths} para uma abordagem detalhada sobre transformações de calibre).
Ao trocarmos $\Phi \left( \mathbf{r},t\right) $ e $\mathbf{A}\left( 
\mathbf{r},t\right) $ nas Eqs. \eqref{eq:field-ele} e \eqref{eq:field-mag} por $\Phi ^{\prime }\left( \mathbf{r},t\right) $ e $%
\mathbf{A}^{\prime }\left( \mathbf{r},t\right) $, dados por%
\begin{equation}
\Phi ^{\prime }\left( \mathbf{r},t\right) =\Phi \left( \mathbf{r},t\right) -%
\frac{\partial \Lambda \left( \mathbf{r},t\right) }{\partial t}  \label{z1}
\end{equation}%
\begin{equation}
\mathbf{A^{\prime }}\left( \mathbf{r},t\right) =\mathbf{A}\left( \mathbf{r}%
,t\right) +\nabla \Lambda \left( \mathbf{r},t\right) \,,  \label{z2}
\end{equation}%
onde $\Lambda \left( \mathbf{r},t\right) $ \'{e} um campo escalar arbitr\'{a}%
rio, os campos elétrico e magnético não são alterados.
Desta forma, temos que resolver a equa\c{c}\~{a}o de Schr\"{o}dinger%
\begin{equation}
i\hbar \frac{\partial \Psi \left( \mathbf{r},t\right) }{\partial t}=\hat{H}%
\Psi \left( \mathbf{r},t\right)
\end{equation}%
onde $\hat{H}$ \'{e} dado pela equa\c{c}\~{a}o (\ref{b4}). Para simplificar
esta equa\c{c}\~{a}o, vamos definir uma nova fun\c{c}\~{a}o de onda%
\begin{equation}
\Psi ^{\prime }\left( \mathbf{r},t\right) =\hat{Q}\left( \mathbf{r},t\right)
\Psi \left( \mathbf{r},t\right)
\end{equation}%
onde introduzimos uma transforma\c{c}\~{a}o unit\'{a}ria%
\begin{equation}
\hat{Q}\left( \mathbf{r},t\right) =\exp \left[ -i\frac{e\Lambda \left( 
\mathbf{r},t\right) }{\hbar }\right] \, ,
\end{equation}%
que satisfaz as rela\c{c}\~{o}es $\hat{Q}^{-1}=\hat{Q}^{\dagger }$ e $\Psi =%
\hat{Q}^{\dagger }\Psi ^{\prime }$. Derivando $\Psi ^{\prime }$ em rela\c{c}%
\~{a}o ao tempo, obtemos%
\begin{equation}
i\hbar \frac{\partial \Psi ^{\prime }\left( \mathbf{r},t\right) }{\partial t}%
=ih\left( \frac{\partial }{\partial t}\hat{Q}\right) \Psi +\hat{Q}\left(
i\hbar \frac{\partial }{\partial t}\Psi \left( \mathbf{r},t\right) \right) =%
\left[ ih\left( \frac{\partial }{\partial t}\hat{Q}\right) \hat{Q}^{\dagger
}+\hat{Q}\hat{H}\hat{Q}^{\dagger }\right] \Psi ^{\prime }=\hat{H}^{\prime
}\Psi ^{\prime }\,,
\end{equation}%
onde 
\begin{equation}
\hat{H}^{\prime }=\frac{\left[ \mathbf{p}+e\mathbf{A}^{\prime }\left( 
\mathbf{r},t\right) \right] ^{2}}{2m}-e\Phi ^{\prime }\left( \mathbf{r}%
,t\right) +V\left( \mathbf{r}\right)
\end{equation}%
e $\Phi ^{\prime }$ e $\mathbf{A}^{\prime }$ s\~{a}o dados pelas equa\c{c}%
\~{o}es (\ref{z1}) e (\ref{z2}).

Lembramos que, no v\'{a}cuo, o campo eletromagn\'{e}tico satisfaz as equa%
\c{c}\~{o}es de Maxwel\cite{jackson}
l%
\begin{eqnarray}
\nabla \cdot \mathbf{D} &=&\rho  \label{ga} \\
\nabla \cdot \mathbf{B} &=&0 \\
\nabla \times \mathbf{E} &=&-\frac{\partial \mathbf{B}}{\partial t}
\label{ga3} \\
\nabla \times \mathbf{H} &=&\frac{\partial \mathbf{D}}{\partial t}+\mathbf{%
j\,},  \label{ga4}
\end{eqnarray}%
onde $\mathbf{D}=\varepsilon _{0}\mathbf{E}$ \'{e} o vetor deslocamento el%
\'{e}trico, $\mathbf{H}=\mu _{0}^{-1}\mathbf{B}$ \'{e} a intensidade do
campo magn\'{e}tico, $\rho $ \'{e} a densidade de carga el\'{e}trica livre, $%
\mathbf{j}$ \'{e} a densidade de corrente el\'{e}trica e $\mu _{0}$ \'{e} a
permeabilidade magn\'{e}tica do v\'{a}cuo. Escolhendo o calibre em que $%
\nabla \cdot \mathbf{A}=0$, chamado de \emph{Calibre de Coulomb}, na aus\^{e}%
ncia de cargas livres a equa\c{c}\~{a}o (\ref{ga}) fornece $\nabla ^{2}\Phi
=0$, o que leva \`{a} solu\c{c}\~{a}o $\Phi =0$. Neste caso, ap\'{o}s
algumas manipula\c{c}\~{o}es usando as propriedades do operador $\nabla $,
obt\'{e}m-se \cite{schleich}%
\begin{equation}
\hat{H}^{\prime }=\frac{\left[ \mathbf{\hat{p}}+e\left( \mathbf{A}+\nabla
\Lambda \right) \right] ^{2}}{2m}+e\frac{\partial \Lambda }{\partial t}%
+V\left( \mathbf{r}\right) \,.
\end{equation}

Se as fontes do campo eletromagn\'{e}tico est\~{a}o longe do \'{a}tomo, das
equa\c{c}\~{o}es (\ref{ga3}) -- (\ref{ga4}) obtemos%
\begin{equation}
\nabla ^{2}\mathbf{A}-\frac{1}{c^{2}}\frac{\partial ^{2}\mathbf{A}}{\partial
t^{2}}=0\,,
\end{equation}%
onde $c=\left( \varepsilon _{0}\mu _{0}\right) ^{-1/2}$ \'{e} a velocidade
da luz no v\'{a}cuo. Numa cavidade unidimensional, com dois espelhos planos
perpendiculares ao eixo $x$, as solu\c{c}\~{o}es s\~{a}o do tipo%
\begin{equation}
\mathbf{A}\left( \mathbf{r},t\right) =\mathbf{A}_{0}\cos \left( kx\right)
\sin \left( \omega t\right) \,,
\end{equation}%
onde $k=2\pi /\lambda $ \'{e} o n\'{u}mero de onda, $\lambda $ \'{e} o
comprimento de onda, $\omega =ck$ \'{e} a frequ\^{e}ncia angular da onda
eletromagn\'{e}tica e $\mathbf{A}_{0}$ \'{e} a amplitude do potencial
vetorial (que depende da sua intensidade, no caso cl\'{a}ssico).

Se o tamanho do \'{a}tomo (denotado por $\Delta r$) for pequeno comparado ao
comprimento de onda da luz, $k\Delta r=2\pi \Delta r/\lambda \ll 1$, podemos
considerar que o potencial vetorial $\mathbf{A}\left( \mathbf{r},t\right) $ 
\'{e} homog\^{e}neo ao longo do \'{a}tomo, de modo que 
\begin{equation}
\mathbf{A}\left( \mathbf{r},t\right) \approx \mathbf{A}\left( t\right) =%
\mathbf{A}_{\ast }\sin \omega t\,\,,
\end{equation}%
onde $\mathbf{A}_{\ast }$ \'{e} a amplitude do potencial vetorial na posi%
\c{c}\~{a}o do \'{a}tomo. Esta \'{e} a chamada \textquotedblleft \emph{%
aproxima\c{c}\~{a}o de dipolo}\textquotedblright . Finalmente, vamos
escolher o campo escalar $\Lambda \left( \mathbf{r},t\right) $ como%
\begin{equation}
\Lambda \left( \mathbf{r},t\right) =-\mathbf{A}\left( \mathbf{r},t\right)
\cdot \mathbf{r}\approx -\mathbf{A}\left( t\right) \cdot \mathbf{r\,},
\end{equation}%
de modo que 
\begin{eqnarray}
\nabla \Lambda \left( \mathbf{r},t\right) &=&-\mathbf{A}\left( t\right) \\
\frac{\partial \Lambda \left( \mathbf{r},t\right) }{\partial t} &=&-\frac{%
\partial \mathbf{A}\left( t\right) }{\partial t}\cdot \mathbf{r=E}\left(
t\right) \cdot \mathbf{r\,}.
\end{eqnarray}

Assim, o Hamiltoniano que descreve a intera\c{c}\~{a}o entre o campo
eletromagn\'{e}tico e o \'{a}tomo, na aproxima\c{c}\~{a}o de dipolo, torna-se%
\begin{equation}
\hat{H}^{\prime }=\frac{\left[ \mathbf{\hat{p}}+e\left( \mathbf{A}-\mathbf{A}%
\right) \right] ^{2}}{2m}+e\mathbf{E}\left( t\right) \cdot \mathbf{r}%
+V\left( \mathbf{r}\right) =\frac{\mathbf{\hat{p}}^{2}}{2m}+V\left( \mathbf{r%
}\right) +e\mathbf{r}\cdot \mathbf{E}\left( t\right) \,.
\end{equation}%
Colocando a origem do sistema de coordenadas sobre o n\'{u}cleo, $\mathbf{r}$
\'{e} a posi\c{c}\~{a}o do el\'{e}tron em rela\c{c}\~{a}o ao n\'{u}cleo.
Portanto, o momento de dipolo el\'{e}trico do \'{a}tomo \'{e} $\mathbf{d}%
=e\left( -\mathbf{r}\right) $. Lembrando a defini\c{c}\~{a}o do Hamiltoniano
de \'{a}tomo isolado, equa\c{c}\~{a}o (\ref{h0}), obtemos na representa\c{c}%
\~{a}o de posi\c{c}\~{a}o%
\begin{equation}
\hat{H}^{\prime }=\hat{H}_{0}-\mathbf{d}\cdot \mathbf{E}\left( t\right) \,.
\end{equation}%
Se n\~{a}o quisermos trabalhar na representa\c{c}\~{a}o de posi\c{c}\~{a}o,
podemos reescrever este Hamiltoniano na forma gen\'{e}rica como%
\begin{equation}
\hat{H}^{\prime }=\hat{H}_{0}-\mathbf{\hat{d}}\cdot \mathbf{\hat{E}}\left(
t\right) \,,  \label{po}
\end{equation}%
onde $\mathbf{\hat{d}}=-e\mathbf{\hat{r}}$ \'{e} o operador dipolo el\'{e}%
trico do \'{a}tomo. Note que esta dedu\c{c}\~{a}o vale para ambas as descri%
\c{c}\~{o}es, cl\'{a}ssica e qu\^{a}ntica, do campo eletromagn\'{e}tico. Por
isso, denotamos o campo el\'{e}trico por operador $\mathbf{\hat{E}}\left(
t\right) $, que no caso semicl\'{a}ssico corresponde a um campo vetorial
tradicional.

Sempre podemos encontrar (anal\'{\i}tica ou numericamente) os autoestados do
Hamiltoniano do \'{a}tomo isolado, resolvendo a equa\c{c}\~{a}o%
\begin{equation}
\hat{H}_{0}|\psi _{n}\rangle =E_{n}|\psi _{n}\rangle \,.
\end{equation}%
Se o \'{a}tomo possui apenas dois n\'{\i}veis de energia, ou se a din\^{a}%
mica do \'{a}tomo est\'{a} confinada a apenas dois n\'{\i}veis de energia,
com as autoenergias $E_{0}$ e $E_{1}$ e os respectivos autoestados $|\psi
_{0}\rangle $ e $|\psi _{1}\rangle $, podemos escrever o operador-identidade
do \'{a}tomo como \cite{sakurai}
\begin{equation}
\hat{1}=|\psi _{0}\rangle \langle \psi _{0}|+|\psi _{1}\rangle \langle \psi
_{1}|\,.
\end{equation}%
Inserindo os operadores-identidade no Hamiltoniano (\ref{po}), obtemos%
\begin{eqnarray}
\hat{H}^{\prime } &=&\hat{H}_{0}\hat{1}-\hat{1}\mathbf{\hat{d}}\hat{1}\cdot 
\mathbf{\hat{E}}\left( t\right)  \notag \\
&=&E_{0}|\psi _{0}\rangle \langle \psi _{0}|+E_{1}|\psi _{1}\rangle \langle
\psi _{1}|-\left( |\psi _{0}\rangle \langle \psi _{0}|+|\psi _{1}\rangle
\langle \psi _{1}|\right) \mathbf{\hat{d}}\left( |\psi _{0}\rangle \langle
\psi _{0}|+|\psi _{1}\rangle \langle \psi _{1}|\right) \cdot \mathbf{\hat{E}}%
\left( t\right) \,.  \label{vi}
\end{eqnarray}

Se, na representa\c{c}\~{a}o de posi\c{c}\~{a}o, as autofun\c{c}\~{o}es do 
\'{a}tomo t\^{e}m paridade definida, isto \'{e}, $\psi _{n}\left( -\mathbf{r}%
\right) =\pm \psi _{n}\left( \mathbf{r}\right) $, teremos%
\begin{equation}
\langle \psi _{n}|\mathbf{\hat{d}}|\psi _{n}\rangle =-e\int \mathbf{r}%
\left\vert \psi _{n}\left( \mathbf{r}\right) \right\vert ^{2}d^{3}r\,,
\end{equation}%
onde a integral \'{e} sobre todo o volume. Como $\mathbf{r}$ \'{e} fun\c{c}%
\~{a}o anti-sim\'{e}trica e $\left\vert \psi _{n}\left( \mathbf{r}\right)
\right\vert ^{2}$ \'{e} sim\'{e}trica, obtemos $\langle \psi _{n}|\mathbf{%
\hat{d}}|\psi _{n}\rangle =0$. Somando e subtraindo o termo constante $%
\left( E_{1}+E_{0}\right) /2$, o Hamiltoniano (\ref{vi}) vira%
\begin{equation}
\hat{H}^{\prime }=\frac{E_{1}+E_{0}}{2}+\left( \frac{E_{1}-E_{0}}{2}\right)
\left( |\psi _{1}\rangle \langle \psi _{1}|-|\psi _{0}\rangle \langle \psi
_{0}|\right) -\left( \mathbf{D}|\psi _{0}\rangle \langle \psi _{1}|+\mathbf{D%
}^{\ast }|\psi _{1}\rangle \langle \psi _{0}|\right) \cdot \mathbf{\hat{E}}%
\left( t\right) \,,
\end{equation}%
onde definimos o elemento n\~{a}o-diagonal da matriz $\mathbf{D}=\langle
\psi _{0}|\mathbf{\hat{d}}|\psi _{1}\rangle $. Como um termo constante n\~{a}%
o afeta a din\^{a}mica, definindo a \emph{frequ\^{e}ncia de transi\c{c}\~{a}%
o at\^{o}mica} $\Omega =\left( E_{1}-E_{0}\right) /\hbar $, podemos escrever
o Hamiltoniano como%
\begin{equation}
\hat{H}^{\prime }=\frac{\hbar \Omega }{2}\hat{\sigma}_{z}-\left( \mathbf{%
\hat{D}}\sigma _{-}+\mathbf{D}^{\ast }\hat{\sigma}_{+}\right) \cdot \mathbf{%
\hat{E}}\left( t\right) \,,
\end{equation}%
onde $\hat{\sigma}_{z}=|\psi _{1}\rangle \langle \psi _{1}|-|\psi
_{0}\rangle \langle \psi _{0}|$, $\hat{\sigma}_{+}=|\psi _{1}\rangle \langle
\psi _{0}|$ e $\hat{\sigma}_{-}=|\psi _{0}\rangle \langle \psi _{1}|$ s\~{a}%
o os famosos operadores de Pauli (ou matrizes de Pauli) para um sistema de
dois n\'{\i}veis.

\begin{figure}[tbh]
\begin{center}
\includegraphics[width=0.29\textwidth]{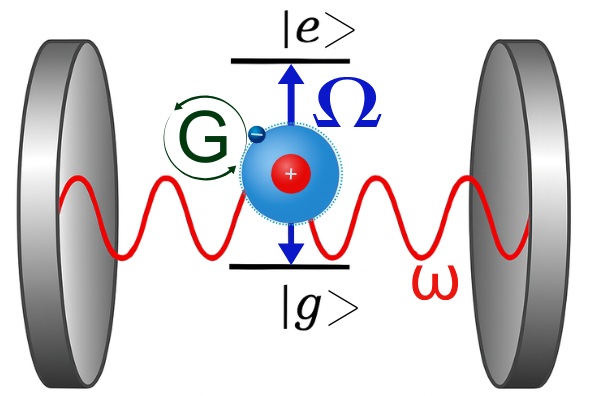} {}
\end{center}
\caption{Vis\~{a}o art\'{\i}stica do modelo de Rabi semicl\'{a}ssico. Um 
\'{a}tomo de dois n\'{\i}veis, com a frequ\^{e}ncia de transi\c{c}\~{a}o $%
\Omega $, interage com um modo do campo eletromagn\'{e}tico de frequ\^{e}%
ncia $\protect\omega $ numa cavidade ideal. $G$ \'{e} a constante de
acoplamento \'{a}tomo-campo devido \`{a} intera\c{c}\~{a}o do tipo dipolo el%
\'{e}trico.}
\label{fig1}
\end{figure}

Se o campo eletromagn\'{e}tico \'{e} tratado classicamente, obtemos%
\begin{equation}
\mathbf{E}\left( t\right) =-\frac{\partial }{\partial t}\left( \mathbf{A}%
_{\ast }\sin \omega t\right) =-\omega \mathbf{A}_{\ast }\cos \omega t\,.
\end{equation}%
Sem perda de generalidade, vamos supor que os estados $|\psi _{0}\rangle $ e 
$|\psi _{1}\rangle $ foram escolhidos de modo a tornar o elemento de matriz $%
\mathbf{D}$ real. Ent\~{a}o, o Hamiltoniano no regime semicl\'{a}ssico vira%
\begin{equation}
\hat{H}^{\prime }=\frac{\hbar \Omega }{2}\hat{\sigma}_{z}+\hbar G\left( \hat{%
\sigma}_{-}+\hat{\sigma}_{+}\right) \cos \omega t \,,  \label{sr}
\end{equation}%
com a constante de acoplamento no regime semicl\'{a}ssico: $G=\left( \mathbf{%
D}\cdot \omega \mathbf{A}_{\ast }\right) /\hbar $. Este \'{e} o \emph{%
Hamiltoniano de Rabi semicl\'{a}ssico}, ilustrado na Figura \ref{fig1}, em
que um \'{a}tomo de dois n\'{\i}veis, tamb\'{e}m chamado de \emph{qubit},
interage com o campo monocrom\'{a}tico cl\'{a}ssico de frequ\^{e}ncia $%
\omega $ dentro de uma cavidade. Usamos a nota\c{c}\~{a}o convencional de 
\'{O}ptica Qu\^{a}ntica, em que o estado de energia mais baixa \'{e} chamado
de \textquotedblleft estado fundamental\textquotedblright\ $|g\rangle $ e o
estado de energia mais alta -- de \textquotedblleft estado
excitado\textquotedblright\ $|e\rangle $ (ou seja, $|\psi _{0}\rangle
=|g\rangle $ e $|\psi _{1}\rangle =|e\rangle $).

Se o campo \'{e} tratado quanticamente, devemos usar o operador campo el\'{e}%
trico, que escrito na representação de Heisenberg é dado por \cite{scully,schleich,knight}%
\begin{equation}
\mathbf{\hat{E}}\left( t\right) =-\mathbf{E}_{0}\left( \hat{a}e^{-i\omega t}+%
\hat{a}^{\dagger }e^{i\omega t}\right) \,,
\end{equation}%
onde $\hat{a}$ e $\hat{a}^{\dagger }$ s\~{a}o os operadores de aniquila\c{c}%
\~{a}o e cria\c{c}\~{a}o para o modo do campo eletromagn\'{e}tico de frequ%
\^{e}ncia $\omega $, que obedecem \`{a} rela\c{c}\~{a}o de comuta\c{c}\~{a}o 
$\left[ \hat{a},\hat{a}^{\dagger }\right] =1$. $\mathbf{E}_{0}$ \'{e} a
amplitude do campo el\'{e}trico na posi\c{c}\~{a}o do \'{a}tomo,
proporcional ao chamado \textquotedblleft campo el\'{e}trico por f\'{o}%
ton\textquotedblright\ ou \textquotedblleft campo el\'{e}trico do v\'{a}%
cuo\textquotedblright 
\begin{equation}
\mathcal{E}=\sqrt{\frac{\hbar \omega }{2\varepsilon _{0}V}}\,,
\end{equation}%
onde $V$ \'{e} o volume da cavidade \cite{schleich,knight}. Neste caso, o
Hamiltoniano vira%
\begin{equation}
\hat{H}^{\prime }=\frac{\hbar \Omega }{2}\hat{\sigma}_{z}+\hbar g\left( \hat{%
\sigma}_{-}+\hat{\sigma}_{+}\right) \left( \hat{a}e^{-i\omega t}+\hat{a}%
^{\dagger }e^{i\omega t}\right) \,,  \label{qr}
\end{equation}%
onde a constante de acoplamento no regime qu\^{a}ntico \'{e} $g=\left( 
\mathbf{D}\cdot \mathbf{E}_{0}\right) /\hbar $. Este \'{e} o \emph{%
Hamiltoniano de Rabi qu\^{a}ntico} (expresso num quadro de intera\c{c}\~{a}o
espec\'{\i}fico, que veremos na seção\ref{QRMz}, em que os operadores de cria\c{c}\~{a}o e aniquila\c{c}\~{a}%
o do campo oscilam com a frequ\^{e}ncia $\omega $). Ressaltamos que a representação da Eq. \eqref{qr} não é usual na literatura.
Geralmente, o Hamiltoniano de Rabi é escrito em termos do termo do átomo (ou qubit), o de campo, e da
interação entre os dois,
\begin{equation}
	\hat{H}^{\prime }=\frac{\hbar \Omega }{2}\hat{\sigma}_{z}+\hbar\omega\hat{n}+\hbar g\left( \hat{%
		\sigma}_{-}+\hat{\sigma}_{+}\right) \left( \hat{a}+\hat{a}%
	^{\dagger }\right) \,,  \label{QRMprime}
\end{equation}
conforme será explicado na seção \ref{QRMz}.

A dedu\c{c}\~{a}o acima foi feita com o intuito de fornecer uma
justificativa minimamente convincente para os Hamiltonianos de Rabi, equa%
\c{c}\~{o}es (\ref{sr}) e (\ref{qr}); para uma dedu\c{c}\~{a}o mais rigorosa
a partir de primeiros princ\'{\i}pios, recomendamos o livro \cite{schleich}.

\section{Modelo de Rabi semicl\'{a}ssico\label{SRM}}

Nesta seção, vamos descrever a dinâmica do modelo de Rabi semiclássico.
Começamos considerando um sistema qu\^{a}ntico de dois n\'{\i}veis, que tamb\'{e}m
chamaremos de qubit ou \'{a}tomo de 2-n\'{\i}veis, descrito pelo
Hamiltoniano (\ref{sr}). Toda a informa\c{c}\~{a}o sobre o \'atomo est\'{a}
contida no seu estado qu\^{a}ntico $|\psi (t)\rangle $, que pode ser escrito
como 
\begin{equation}
|\psi \left( t\right) \rangle =c_{g}\left( t\right) |g\rangle +c_{e}\left(
t\right) |e\rangle ~,
\end{equation}%
onde $t$ \'{e} o tempo, $c_{g}\left( t\right) $ \'{e} a amplitude de
probabilidade do estado fundamental e $c_{e}\left( t\right) $ \'{e} a
amplitude de probabilidade do estado excitado. Na representa\c{c}\~{a}o
matricial, vamos definir%
\begin{equation}
|g\rangle =\left( 
\begin{array}{c}
0 \\ 
1%
\end{array}%
\right) \quad \text{e \quad }|e\rangle =\left( 
\begin{array}{c}
1 \\ 
0%
\end{array}%
\right) \,,
\end{equation}%
de modo que as matrizes de Pauli adquirem a sua forma usual:%
\begin{equation}
\hat{\sigma}_{z}=\left( 
\begin{array}{cc}
1 & 0 \\ 
0 & -1%
\end{array}%
\right) ~,~\hat{\sigma}_{+}=\left( 
\begin{array}{cc}
0 & 1 \\ 
0 & 0%
\end{array}%
\right) ,\text{ }\hat{\sigma}_{-}=\left( 
\begin{array}{cc}
0 & 0 \\ 
1 & 0%
\end{array}%
\right) \,.
\end{equation}

O estado inicial do \'{a}tomo, $|\psi \left( 0\right) \rangle $, e o
Hamiltoniano definem completamente a evolu\c{c}\~{a}o temporal do sistema
atrav\'{e}s da Equa\c{c}\~{a}o de Schr\"{o}dinger%
\begin{equation}
i\hbar \frac{\partial |\psi (t)\rangle }{\partial t}=\hat{H}^{\prime }|\psi
(t)\rangle \,.
\end{equation}%
Na representa\c{c}\~{a}o matricial, ela torna-se%
\begin{equation}
i\hbar \left( 
\begin{array}{c}
\frac{dc_{e}\left( t\right) }{dt} \\ 
\frac{dc_{g}\left( t\right) }{dt}%
\end{array}%
\right) =\left( 
\begin{array}{cc}
\hbar \Omega /2 & \hbar G\cos \omega t \\ 
\hbar G\cos \omega t & -\hbar \Omega /2%
\end{array}%
\right) \left( 
\begin{array}{c}
c_{e}\left( t\right) \\ 
c_{g}\left( t\right)%
\end{array}%
\right) \,\,,
\end{equation}%
e a nossa tarefa consiste em resolver o par de equa\c{c}\~{o}es diferenciais
ordin\'{a}rias acopladas%
\begin{eqnarray}
i\dot{c}_{e} &=&\frac{\Omega }{2}c_{e}+G\cos \left( \omega t\right) c_{g}
\label{f1} \\
i\dot{c}_{g} &=&-\frac{\Omega }{2}c_{g}+G\cos \left( \omega t\right) c_{e}\,.
\label{f2}
\end{eqnarray}

Para resolv\^{e}-las, primeiro vamos definir as novas amplitudes de
probabilidade%
\begin{equation}
C_{e}=c_{e}e^{i\omega t/2}\quad ,\quad C_{g}=c_{g}e^{-i\omega t/2}\,.
\end{equation}%
Substituindo-as nas equa\c{c}\~{o}es (\ref{f1}) -- (\ref{f2}) e escrevendo $%
\cos \omega t=\left( e^{i\omega t}+e^{-i\omega t}\right) /2$, obtemos%
\begin{eqnarray}
i\dot{C}_{e} &=&\left( \frac{\Delta }{2}C_{e}+\frac{G}{2}C_{g}\right) +\frac{%
G}{2}e^{2i\omega t}C_{g}  \label{i1} \\
i\dot{C}_{g} &=&\left( -\frac{\Delta }{2}C_{g}+\frac{G}{2}C_{e}\right) +%
\frac{G}{2}e^{-2i\omega t}C_{e}\,,  \label{i2}
\end{eqnarray}%
onde definimos o \textquotedblleft \emph{detuning}\textquotedblright\
(dessintonia) entre as frequ\^{e}ncias do \'{a}tomo e do campo eletromagn%
\'{e}tico 
\begin{equation}
\Delta =\Omega -\omega .
\end{equation}
Quando este parâmetro for nulo, o campo estará em ressonância com o átomo.

\subsection{M\'{e}todo aproximado para a solu\c{c}\~{a}o anal\'{\i}tica}

Existem v\'{a}rias maneiras de resolver aproximadamente as equa\c{c}\~{o}es (%
\ref{i1}) -- (\ref{i2}) \cite{graham,munz,Liu,Lu,merlin,castanos}. Aqui,
vamos descrever uma abordagem que envolve somente algumas substitui\c{c}\~{o}%
es e derivadas simples. Com base nas refer\^{e}ncias \cite{marinho2,acosta1}%
, vamos definir novas fun\c{c}\~{o}es temporais $A_{\pm }(t)$:%
\begin{equation}
C_{g}\left( t\right) =\frac{G}{2\sqrt{R}}\left[ e^{-iRt/2}\sqrt{\frac{1}{%
R_{-}}}A_{+}(t)+e^{iRt/2}\sqrt{\frac{1}{R_{+}}}A_{-}(t)\right]  \label{n1}
\end{equation}%
\begin{equation}
C_{e}\left( t\right) =\frac{G}{2\sqrt{R}}\left[ e^{-iRt/2}\sqrt{\frac{1}{%
R_{+}}}A_{+}(t)-e^{iRt/2}\sqrt{\frac{1}{R_{-}}}A_{-}(t)\right]\,,  \label{n2}
\end{equation}%
onde $R=\sqrt{G^{2}+\Delta ^{2}}$ e $R_{\pm }=\left( R\pm \Delta \right) /2$%
. Essas substitui\c{c}\~{o}es, embora pare\c{c}am complicadas e
contraintuitivas, surgem naturalmente ao diagonalizarmos uma matriz $2\times
2$, mas aqui n\~{a}o vamos detalhar este processo. Uma vez determinadas as
fun\c{c}\~{o}es $A_{\pm }(t)$, a probabilidade de encontrar o \'{a}tomo no
estado excitado fica dada por%
\begin{equation}
P_{e}(t)=\left\vert c_{e}(t)\right\vert ^{2}=\left\vert C_{e}(t)\right\vert
^{2}=\frac{G^{2}}{4R}\left\vert e^{-iRt/2}A_{+}(t)\sqrt{\frac{1}{R_{-}}}%
-e^{iRt/2}A_{-}(t)\sqrt{\frac{1}{R_{+}}}\right\vert ^{2}\,.  \label{prob}
\end{equation}

Invertendo as equa\c{c}\~{o}es (\ref{n1}) -- (\ref{n2}), obtemos%
\begin{equation}
A_{+}(t)=\left[ \sqrt{\frac{R_{-}}{R}}C_{g}(t)+\sqrt{\frac{R_{+}}{R}}C_{e}(t)%
\right] e^{iRt/2}
\end{equation}%
\begin{equation}
A_{-}(t)=\left[ \sqrt{\frac{R_{+}}{R}}C_{g}(t)-\sqrt{\frac{R_{-}}{R}}C_{e}(t)%
\right] e^{-iRt/2}\,.
\end{equation}%
Derivando estas equa\c{c}\~{o}es em rela\c{c}\~{a}o ao tempo e usando as equa%
\c{c}\~{o}es originais (\ref{i1}) e (\ref{i2}), depois de algumas manipula%
\c{c}\~{o}es alg\'{e}bricas, encontramos as equa\c{c}\~{o}es%
\begin{equation}
i\dot{A}_{+}(t)=\frac{G^{2}}{2R}\cos \left( 2\omega t\right) A_{+}(t)+\frac{G%
}{2R}\left[ R_{+}e^{2i\omega t}-R_{-}e^{-2i\omega t}\right] e^{iRt}A_{-}(t)
\label{d1}
\end{equation}%
\begin{equation}
i\dot{A}_{-}(t)=-\frac{G^{2}}{2R}\cos \left( 2\omega t\right) A_{-}(t)+\frac{G}{2R}%
\left[ R_{+}e^{-2i\omega t}-R_{-}e^{2i\omega t}\right] e^{-iRt}A_{+}(t)\,.
\label{d2}
\end{equation}

A solu\c{c}\~{a}o aproximada mais simples do modelo de Rabi consiste em nada
mais que desprezar o lado direito das equa\c{c}\~{o}es (\ref{d1}) -- (\ref%
{d2}), de modo que as fun\c{c}\~{o}es $A_{\pm }\left( t\right) $ n\~{a}o
variam no tempo, sendo iguais aos seus valores iniciais $A_{\pm }\left(
0\right) $. Esta \'{e} a famosa \textquotedblleft Aproxima\c{c}\~{a}o de
Onda Girante\textquotedblright , mais conhecida pelo termo em ingl\^{e}s
\textquotedblleft \emph{Rotating Wave Approximation}\textquotedblright\
(RWA), adotada em muitos livros-texto de \'{O}ptica Qu\^{a}ntica e Informa%
\c{c}\~{a}o Qu\^{a}ntica \cite{boyd,knight,scully}; ela consiste em
desprezar completamente os termos \textquotedblleft rapidamente oscilantes
no tempo\textquotedblright, com as frequ\^{e}ncias $2\omega $ e $2\omega \pm
R$. Se o \'{a}tomo estava inicialmente no estado fundamental, as condi\c{c}%
\~{o}es iniciais eram $c_{g}(0)=C_{g}(0)=1$ e $c_{e}(0)=C_{e}(0)=0$. Da equa%
\c{c}\~{a}o (\ref{prob}) encontramos, sob a aproxima\c{c}\~{a}o RWA:%
\begin{equation}
P_{e}^{(RWA)}(t)=\frac{G^{2}}{4R^{2}}\left\vert
e^{-iRt/2}-e^{iRt/2}\right\vert ^{2}=\frac{G^{2}}{R^{2}}\sin ^{2}\left( 
\frac{Rt}{2}\right) \,.  \label{rwa}
\end{equation}%
Em particular, para a dessintonia nula, $\Delta =0$, obtemos $%
P_{e}^{(RWA)}(t)=\sin ^{2}\left( Gt/2\right) $. Por isso, a dura\c{c}\~{a}o
do chamado \textquotedblleft pulso-$\pi $\textquotedblright, para o qual o 
\'{a}tomo sofre a transi\c{c}\~{a}o completa do estado $|g\rangle $ para o
estado $|e\rangle $, \'{e} $T_{\pi }=\pi /G$. \`{A} medida que a dessintonia
aumenta, as oscila\c{c}\~{o}es ficam mais r\'{a}pidas, por\'{e}m a sua
amplitude diminui para $\left[ 1+\left( \Delta /G\right) ^{2}\right] ^{-1}$.

Entretanto, a aproxima\c{c}\~{a}o RWA abre m\~ao de diversos fen\^{o}menos f%
\'{\i}sicos importantes, como as resson\^{a}ncias multifot\^{o}nicas e caoticidade \cite{belebrov, milonni}.
Felizmente, existe um jeito relativamente simples de encontrar solu\c{c}\~{o}%
es aproximadas bastante precisas usando uma simples substitui\c{c}\~{a}o:%
\begin{equation}
A_{\pm }(t)=e^{\mp i\Upsilon \sin \left( 2\omega t\right) /2}a_{\pm }(t)\,~,
\label{a2}
\end{equation}%
com a condi\c{c}\~{a}o inicial $A_{\pm }(0)=a_{\pm }(0)$, e definimos o par%
\^{a}metro adimensional%
\begin{equation}
\Upsilon \equiv \frac{G^{2}}{2\omega R}\,\,.
\end{equation}%
Com estas defini\c{c}\~{o}es, a probabilidade de encontrar o \'{a}tomo no
estado excitado passa a ser%
\begin{equation}
P_{e}(t)=\frac{1}{2}\left\vert e^{-i\left( \Upsilon \sin 2\omega t+Gt\right)
}a_{+}(t)-a_{-}(t)\right\vert ^{2}\,\,.  \label{sa}
\end{equation}%
Por exemplo, para a dessintonia nula, temos $R=G$, $R_{\pm }=G/2$ e $%
\Upsilon \equiv G/(2\omega )$; por outro lado, para $\Delta \gg G$ (quando
ocorrem as resson\^{a}ncias multifot\^{o}nicas) temos $R\approx \Delta
+G^{2}/(2\Delta ) $, $R_{-}\approx G^{2}/(4\Delta )$, $R_{+}\approx \Delta
+G^{2}/(4\Delta )$ e $\Upsilon \approx G^{2}/(2\omega \Delta )$.

Substituindo a equa\c{c}\~{a}o (\ref{a2}) nas equa\c{c}\~{o}es (\ref{d1}) --
(\ref{d2}), obtemos equa\c{c}\~{o}es simples para $a_{\pm }$%
\begin{equation}
\dot{a}_{+}=-iQ_{t}a_{-}\quad ,\quad \dot{a}_{-}=-iQ_{t}^{\ast }a_{+}~,
\label{apm}
\end{equation}%
onde%
\begin{equation}
Q_{t}=\frac{G}{2R}e^{iRt}e^{i\Upsilon \cos \left( 2\omega t-\pi /2\right)
}\left( R_{+}e^{2i\omega t}-R_{-}e^{-2i\omega t}\right) \,  \label{qtr}
\end{equation}%
e $Q_{t}^{\ast }$ denota o complexo conjugado de $Q_{t}$. Agora, vamos
invocar a expans\~{a}o de Jacobi-Anger, bem conhecida dos livros-texto de F%
\'{\i}sica Matem\'{a}tica \cite{arfken}:%
\begin{equation}
e^{i\Upsilon \cos \theta }=J_{0}\left( \Upsilon \right) +\sum_{n=1}^{\infty
}i^{n}J_{n}\left( \Upsilon \right) \left( e^{i\theta n}+e^{-i\theta
n}\right) ~,  \label{anger}
\end{equation}%
onde 
\begin{equation}
J_{n}\left( \Upsilon \right) =\sum_{k=0}^{\infty }\frac{\left( -1\right) ^{k}%
}{k!\left( n+k\right) !}\left( \frac{\Upsilon }{2}\right) ^{n+2k}
\label{pan}
\end{equation}%
\'{e} a fun\c{c}\~{a}o de Bessel do primeiro tipo, que possui uma
propriedade \'{u}til: $J_{n}\left( -x\right) =\left( -1\right)
^{n}J_{n}\left( x\right) $. Felizmente, as fun\c{c}\~{o}es de Bessel s\~{a}o
tabeladas ou podem ser facilmente calculadas usando softwares cient\'{\i}%
ficos (como Maple, Mathematica, Fortran, etc), ent\~{a}o, neste trabalho, n%
\~{a}o vamos precisar calcular as fun\c{c}\~{o}es de Bessel usando a expans%
\~{a}o (\ref{pan}). Para encurtar a nota\c{c}\~{a}o, daqui para frente vamos
omitir o argumento $\Upsilon $ das fun\c{c}\~{o}es de Bessel, tendo em mente
que $J_{n}\equiv J_{n}\left( \Upsilon \right) $.

\subsection{Resson\^{a}ncia de um f\'{o}ton para $\Omega =\protect\omega $}

Primeiro, vamos analisar o caso de dessintonia nula e a situa\c{c}\~{a}o
usual em que $G\ll 2\omega $, de modo que $\Upsilon \ll 1$. Mantendo apenas
os termos proporcionais a $J_{n}\left( \Upsilon \right) $, com $n\leq 2$ na
expans\~{a}o (\ref{anger}), a fun\c{c}\~{a}o $Q_{t}$ torna-se%
\begin{equation}
Q_{t}\approx i\frac{G}{2}\left[ \left( J_{0}-J_{2}\right) \sin 2\omega
t+iJ_{1}\right] e^{iGt}\,\,.  \label{qt}
\end{equation}%
Uma solu\c{c}\~{a}o um pouco mais precisa que a equa\c{c}\~{a}o (\ref{rwa}) 
\'{e} obtida desprezando o termo proporcional a $\sin 2\omega t$. Neste
caso, as equa\c{c}\~{o}es diferenciais passam a ser%
\begin{equation}
\dot{a}_{\pm }=i\frac{GJ_{1}}{2}e^{\pm iGt}a_{\mp }~
\end{equation}%
e podem ser facilmente transformadas em equa\c{c}\~{o}es diferenciais ordin%
\'{a}rias de segunda ordem com coeficientes constantes:%
\begin{equation}
\ddot{a}_{\pm }\mp iG\dot{a}_{\pm }+\left( \frac{GJ_{1}}{2}\right)
^{2}a_{\pm }=0\,.
\end{equation}%
As solu\c{c}\~{o}es s\~{a}o%
\begin{equation}
a_{+}\left( t\right) =e^{iGt/2}\left[ b_{1}e^{iGst/2}+b_{2}e^{-iGst/2}\right]
\label{w1}
\end{equation}%
\begin{equation}
a_{-}\left( t\right) =\frac{e^{-iGt/2}}{J_{1}}\left[ b_{1}\left( 1+s\right)
e^{iGst/2}+b_{2}\left( 1-s\right) e^{-iGst/2}\right] \,,  \label{w2}
\end{equation}%
onde $s=\sqrt{1+J_{1}^{2}}$ e os coeficientes $b_{1}$ e $b_{2}$ dependem das
condi\c{c}\~{o}es iniciais:%
\begin{eqnarray}
b_{1} &=&\frac{\left( s-1+J_{1}\right) c_{g}\left( 0\right) +\left(
s-1-J_{1}\right) c_{e}\left( 0\right) }{2\sqrt{2}s} \\
b_{2} &=&\frac{\left( s+1-J_{1}\right) c_{g}\left( 0\right) +\left(
s+1+J_{1}\right) c_{e}\left( 0\right) }{2\sqrt{2}s}.
\end{eqnarray}%
Chamamos esta solu\c{c}\~{a}o de \textquotedblleft solu\c{c}\~{a}o intermedi%
\'{a}ria\textquotedblright , pois daqui a pouco veremos que ela se aproxima
mais da solu\c{c}\~{a}o num\'{e}rica exata do que a solu\c{c}\~{a}o RWA.

Uma solu\c{c}\~{a}o ainda mais precisa consiste em resolver numericamente as
equa\c{c}\~{o}es (\ref{apm}) com a fun\c{c}\~{a}o $Q_{t}$ dada pela equa\c{c}%
\~{a}o (\ref{qt}). Chamamos esta solu\c{c}\~{a}o de \emph{semianal\'{\i}tica}%
, pois uma vez encontradas as fun\c{c}\~{o}es $a_{\pm }\left( t\right) $, a
probabilidade de excita\c{c}\~{a}o do \'{a}tomo \'{e} dada pela f\'{o}rmula (%
\ref{sa}).

\begin{figure}[tbh]
\begin{center}
\includegraphics[width=0.99\textwidth]{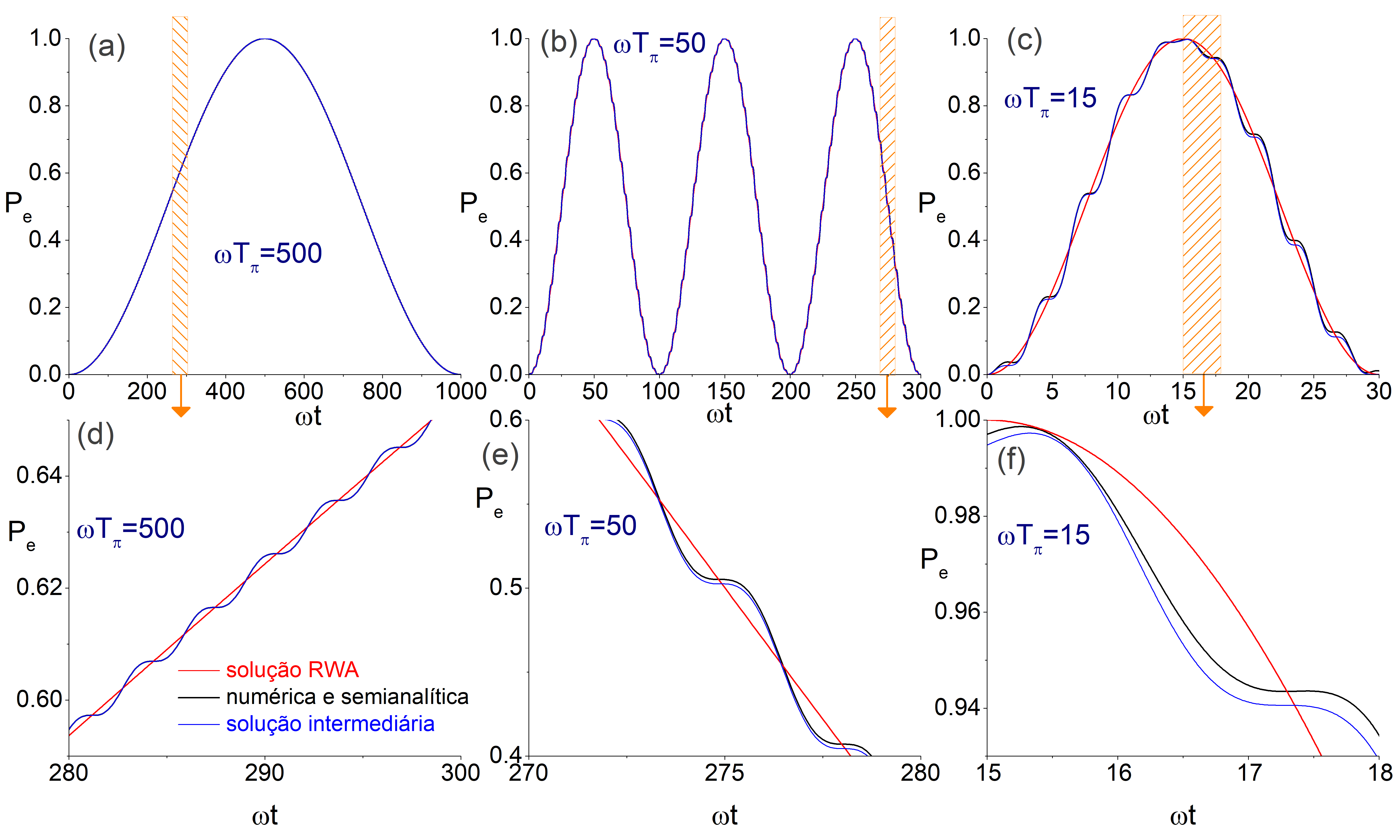} {}
\end{center}
\caption{Comportamento da probabilidade de excita\c{c}\~{a}o do \'{a}tomo, $%
P_{e}(t)$, em fun\c{c}\~{a}o do tempo adimensional $\protect\omega t$, para
o estado inicial $|g\rangle $ e diferentes valores da constante de
acoplamento $G$. Em vez de especificar $G$, indicamos a dura\c{c}\~{a}o do
pulso-$\protect\pi $, dado por $\protect\omega T_{\protect\pi }=\protect\pi 
\protect\omega /G$. As figuras na linha de baixo s\~{a}o zooms das figuras
na linha de cima para os intervalos de tempo indicados por ret\^{a}ngulos
hachurados. Esta figura ilustra bem a precis\~{a}o das solu\c{c}\~{o}es
aproximadas para diferentes valores de $T_{\protect\pi }$.}
\label{fig2}
\end{figure}

O comportamento da probabilidade de excita\c{c}\~{a}o at\^{o}mica em fun\c{c}%
\~{a}o do tempo, $P_{e}(t)$, est\'{a} ilustrado na Figura \ref{fig2} para a
dessintonia nula. Consideramos o \'{a}tomo inicialmente no estado
fundamental, $|\psi \left( 0\right) \rangle =|g\rangle $, e escolhemos tr%
\^{e}s valores diferentes da constante de acoplamento $G$. Em vez de indicar
o valor de $G$, preferimos indicar o valor da dura\c{c}\~{a}o do pulso-$\pi $%
, relacionado com o valor de $G$ pela rela\c{c}\~{a}o $G=\pi /T_{\pi }$: na
figura \ref{fig2}a $\omega T_{\pi }=500$, na figura \ref{fig2}b $\omega
T_{\pi }=50$ e na figura \ref{fig2}c $\omega T_{\pi }=15$. As figuras \ref%
{fig2}d, \ref{fig2}e e \ref{fig2}f s\~{a}o zooms das figuras \ref{fig2}a, %
\ref{fig2}b e \ref{fig2}c, respectivamente, para os intervalos de tempo
indicados por ret\^{a}ngulos hachurados. Linhas pretas mostram a solu\c{c}%
\~{a}o semianal\'{\i}tica, indistingu\'{\i}vel da solu\c{c}\~{a}o num\'{e}%
rica exata das equa\c{c}\~{o}es (\ref{f1}) -- (\ref{f2}) para todos os casos
considerados neste trabalho; linhas vermelhas mostram a solu\c{c}\~{a}o padr%
\~{a}o de livros-texto sob a aproxima\c{c}\~{a}o RWA, equa\c{c}\~{a}o (\ref%
{rwa}); linhas azuis mostram a solu\c{c}\~{a}o intermedi\'{a}ria, equa\c{c}%
\~{o}es (\ref{w1}) -- (\ref{w2}). Vemos que para $\omega T_{\pi }=500$
(pulsos \textquotedblleft longos\textquotedblright ), a solu\c{c}\~{a}o
intermedi\'{a}ria coincide com a semianal\'{\i}tica, enquanto a solu\c{c}%
\~{a}o RWA n\~{a}o exibe as pequenas oscila\c{c}\~{o}es de $P_{e}(t)$
previstas pelo Hamiltoniano de Rabi. Para $\omega T_{\pi }=50$ a solu\c{c}%
\~{a}o intermedi\'{a}ria come\c{c}a a destoar da solu\c{c}\~{a}o semianal%
\'{\i}tica depois de algumas oscila\c{c}\~{o}es de Rabi ($t\gtrsim 5T_{\pi }$%
), embora ainda seja bastante precisa para intervalos de tempo menores. Por 
\'{u}ltimo, para $\omega T_{\pi }=15$, tanto a solu\c{c}\~{a}o RWA quanto a
intermedi\'{a}ria diferem da solu\c{c}\~{a}o semianal\'{\i}tica desde o come%
\c{c}o.

\begin{figure}[tbh]
\begin{center}
\includegraphics[width=0.99\textwidth]{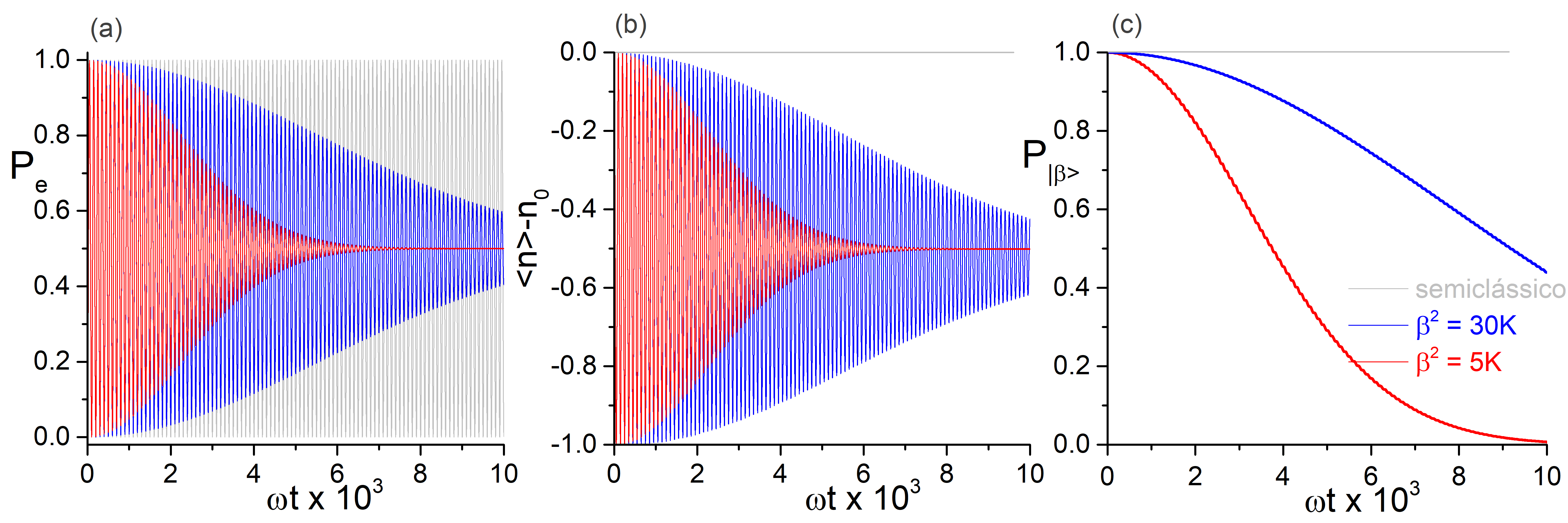} {}
\end{center}
\caption{Compara\c{c}\~{a}o de modelos de Rabi semicl\'{a}ssico e qu\^{a}%
ntico para $\protect\omega T_{\protect\pi }=50$ e $\Delta =0$. a)
Probabilidade de excita\c{c}\~{a}o do \'{a}tomo, partindo do estado inicial $%
|g\rangle $, para o modelo semicl\'{a}ssico (linha cinza) e modelo qu\^{a}%
ntico com $\protect\beta ^{2}=5$K (linha vermelha) e $\protect\beta ^{2}=30$%
K (linha azul), onde $K$ denota $10^{3}$ e $\protect\beta $ \'{e} a
amplitude do estado coerente $|\mathbf{\protect\beta }\rangle $. b) Varia%
\c{c}\~{a}o do n\'{u}mero m\'{e}dio de f\'{o}tons na cavidade em fun\c{c}%
\~{a}o do tempo. c) Probabilidade de que o campo eletromagn\'{e}tico
permanece no estado coerente inicial $|\mathbf{\protect\beta }\rangle $.
Para o modelo semicl\'{a}ssico, acrescentamos artificialmente as linhas
cinzas constantes nas figuras (b) e (c), $\left\langle n\right\rangle
-n_{0}=0$ e $P_{|\protect\beta \rangle }=1$, pois, neste caso, o campo
eletromagn\'{e}tico \'e considerado imut\'avel.}
\label{fig3}
\end{figure}

Assim, vemos que podemos controlar o estado do \'{a}tomo usando o campo
eletromagn\'{e}tico. Por\'{e}m, a f\'{o}rmula anal\'{\i}tica sob a aproxima%
\c{c}\~{a}o RWA, geralmente apresentada nos livros-texto, n\~{a}o consegue
descrever de forma quantitativamente precisa todos os detalhes da din\^{a}%
mica, fornecendo apenas o comportamento m\'{e}dio de $P_{e}(t)$. Para tempos
maiores, o \'{a}tomo exibe oscila\c{c}\~{o}es de Rabi enquanto durar a intera%
\c{c}\~{a}o com o campo eletromagn\'{e}tico \emph{cl\'assico}, conforme
ilustrado pela linha cinza na Figura \ref{fig3}a para $\omega T_{\pi }=50$
(as solu\c{c}\~{o}es semianal\'{\i}tica e a num\'{e}rica exata coincidem
para todos os tempos, neste caso). As demais curvas na Figura \ref{fig3}
correspondem ao modelo de Rabi Qu\^{a}ntico, e ser\~{a}o explicadas na se%
\c{c}\~{a}o \ref{QRMz}.

\subsection{Resson\^{a}ncia de m\'{u}ltiplos f\'{o}tons para $\Omega \approx
\left( 2K+1\right) \protect\omega $}

Agora, vamos considerar o regime $\Delta ,2\omega \gg G$. Usando a expans%
\~{a}o (\ref{anger}), depois de agrupar termos similares, encontramos%
\begin{equation}
Q_{t}=\sum_{n=0}^{\infty }\Lambda _{n}e^{i\left( 2\omega n+R\right)
t}+\sum_{k=1}^{\infty }L_{k}e^{i\left( R-2\omega k\right) t}\,,  \label{oi}
\end{equation}%
onde definimos coeficientes constantes%
\begin{eqnarray}
\Lambda _{0} &=&-\frac{G}{2}J_{1}  \label{m1} \\
\Lambda _{n} &=&\frac{G}{2R}\left( R_{+}J_{n-1}-R_{-}J_{n+1}\right)
\label{m2} \\
L_{k} &=&\left( -1\right) ^{k+1}\frac{G}{2R}\left(
R_{+}J_{k+1}-R_{-}J_{k-1}\right) \,.  \label{m3}
\end{eqnarray}%
No nosso regime de par\^{a}metros, apenas um \'{u}nico termo do tipo $%
L_{K}e^{i\left( R-2\omega K\right) t}$ varia lentamente com o tempo, quando $%
R-2\omega K$ est\'{a} pr\'{o}ximo de zero (onde $K$ \'{e} algum n\'{u}mero
inteiro). Portanto, podemos escrever de forma gen\'{e}rica%
\begin{equation}
Q_{t}=L_{K}e^{i\left( R-2\omega K\right) t}+q_{t}~,  \label{eq}
\end{equation}%
onde $q_{t}$ inclui todos os termos que oscilam rapidamente com tempo na equa%
\c{c}\~{a}o (\ref{oi}). Vamos reescrever $q_{t}$ como%
\begin{equation}
q_{t}=\sum_{i=1}^{\infty }p_{i}e^{itf_{i}}=p_{1}e^{itf_{1}}\left[
1+\sum_{i=2}^{\infty }\frac{p_{i}}{p_{1}}e^{it\left( f_{i}-f_{1}\right) }%
\right] ~,  \label{jo}
\end{equation}%
onde definimos $p_{1}$ como o maior (em m\'{o}dulo) de todos os
coeficientes: $\left\vert p_{1}\right\vert \gg \left\vert p_{j}\right\vert $
para $j>1$. Al\'{e}m disso, para todos os coeficientes temos $\left\vert
p_{i}\right\vert \ll \left\vert f_{i}\right\vert $. Das f\'{o}rmulas (\ref%
{m1}) -- (\ref{m3}) constatamos que $p_{1}=\Lambda _{1}$ e $f_{1}=\left(
2\omega +R\right) $, enquanto $p_{j>1}$ correspondem aos demais coeficientes 
$\Lambda _{j\neq 1}$ e $L_{j\neq K}$.

\subsubsection{Elimina\c{c}\~{a}o dos termos que oscilam rapidamente}

Primeiro, vamos determinar a solu\c{c}\~{a}o quando $L_{K}=0$, de modo que
sobram apenas termos rapidamente oscilantes na equa\c{c}\~{a}o (\ref{eq}).
Podemos facilmente obter as seguintes equa\c{c}\~{o}es diferenciais
desacopladas para as fun\c{c}\~{o}es $a_{\pm }$%
\begin{equation}
\ddot{a}_{+}-\frac{\dot{q}_{t}}{q_{t}}\dot{a}_{+}+\left\vert
q_{t}\right\vert ^{2}a_{+}=0  \label{ap}
\end{equation}%
\begin{equation}
\ddot{a}_{-}-\left( \frac{\dot{q}_{t}}{q_{t}}\right) ^{\ast }\dot{a}%
_{-}+\left\vert q_{t}\right\vert ^{2}a_{-}=0\,.  \label{am}
\end{equation}%
Da equa\c{c}\~{a}o (\ref{jo}) encontramos%
\begin{equation}
\left\vert q_{t}\right\vert ^{2}=\sum_{i=1}^{\infty }p_{i}^{2}+\sum_{i,j\neq
i}^{\infty }p_{i}p_{j}e^{it\left( f_{i}-f_{j}\right) }
\end{equation}%
\begin{equation}
\frac{1}{q_{t}}\approx \frac{1}{p_{1}}e^{-itf_{1}}\left[ 1-\sum_{i=2}^{%
\infty }\frac{p_{i}}{p_{1}}e^{it\left( f_{i}-f_{1}\right) }\right]
\end{equation}%
\begin{equation}
\frac{\dot{q}_{t}}{q_{t}}\approx i\left( f_{1}+\sum_{i=2}^{\infty }\left(
f_{i}-f_{1}\right) \frac{p_{i}}{p_{1}}e^{it\left( f_{i}-f_{1}\right)
}-\sum_{i,j=2}^{\infty }f_{i}\frac{p_{i}}{p_{1}}\frac{p_{j}}{p_{1}}%
e^{it\left( f_{i}+f_{j}-2f_{1}\right) }\right) \,.
\end{equation}

Agora, vamos fazer a \emph{aproxima\c{c}\~{a}o f\'{\i}sica mais importante}
desta se\c{c}\~{a}o: postulamos que, no regime de interesse, as fun\c{c}\~{o}%
es $a_{\pm }$ variam muito pouco durante o intervalo de tempo $\tau =2\pi
/\left\vert f_{1}\right\vert $. Fazendo a m\'{e}dia temporal sobre o per%
\'{\i}odo $\tau $ dos dois lados da equa\c{c}\~{a}o (\ref{ap}), considerando
que $a_{+}$ praticamente n\~{a}o varia neste intervalo, obtemos%
\begin{equation}
\ddot{a}_{+}\left( t\right) -\dot{a}_{+}\left( t\right) \frac{1}{\tau }%
\int_{t}^{t+\tau }dt^{\prime }\frac{\dot{q}_{t}}{q_{t}}+a_{+}\left( t\right) 
\frac{1}{\tau }\int_{t}^{t+\tau }dt^{\prime }\left\vert q_{t}\right\vert
^{2}=0\,.
\end{equation}%
A primeira m\'{e}dia \'{e} dada por%
\begin{equation}
\frac{1}{\tau }\int_{t}^{t+\tau }dt^{\prime }\frac{\dot{q}_{t}}{q_{t}}%
\approx if_{1}+\frac{\left\vert f_{1}\right\vert }{2\pi }\sum_{i=2}^{\infty }%
\frac{p_{i}}{p_{1}}\left[ e^{i\tau f_{i}}-1\right] e^{it\left(
f_{i}-f_{1}\right) }-\frac{\left\vert f_{1}\right\vert }{2\pi }%
\sum_{i,j=2}^{\infty }\frac{f_{i}}{\left( f_{i}+f_{j}-2f_{1}\right) }\frac{%
p_{i}}{p_{1}}\frac{p_{j}}{p_{1}}\left[ e^{i\tau \left( f_{i}+f_{j}\right) }-1%
\right] e^{it\left( f_{i}+f_{j}-2f_{1}\right) }\,.
\end{equation}%
Se $f_{i}+f_{j}=2f_{1}$, temos $e^{i\tau \left( f_{i}+f_{j}\right)
}-1=e^{2i\tau f_{1}}-1=0$. Portanto, vemos que para $\left\vert
p_{i>1}\right\vert \ll \left\vert p_{1}\right\vert $, a express\~{a}o acima
fica aproximadamente igual a $if_{1}$. A outra m\'{e}dia \'{e} facilmente
calculada como 
\begin{equation}
\frac{1}{\tau }\int_{t}^{t+\tau }dt^{\prime }\left\vert q_{t}\right\vert
^{2}=\sum_{i=1}^{\infty }p_{i}^{2}+\frac{\left\vert f_{1}\right\vert }{2\pi }%
\sum_{i,j\neq i}^{\infty }p_{i}p_{j}\frac{e^{i\tau \left( f_{i}-f_{j}\right)
}-1}{i\left( f_{i}-f_{j}\right) }e^{it\left( f_{i}-f_{j}\right) }\approx
\sum_{i=1}^{\infty }p_{i}^{2}\,.
\end{equation}%
Portanto, obtemos%
\begin{equation}
\ddot{a}_{+}-if_{1}\dot{a}_{+}+\sum_{i=1}^{\infty }p_{i}^{2}a_{+}=0
\label{so1}
\end{equation}%
Repetindo a mesma an\'{a}lise para a equa\c{c}\~{a}o (\ref{am}), encontramos%
\begin{equation}
\ddot{a}_{-}+if_{1}\dot{a}_{-}+\sum_{i=1}^{\infty }p_{i}^{2}a_{-}=0
\label{so2}
\end{equation}

As solu\c{c}\~{o}es destas equa\c{c}\~{o}es, que variam lentamente na escala
de tempo $\tau $, s\~{a}o%
\begin{equation}
a_{+}=b_{+}e^{-it\delta }~,~a_{-}=b_{-}e^{it\delta }~,  \label{sol}
\end{equation}%
onde $b_{\pm }$ s\~{a}o constantes, e o pequeno \emph{deslocamento de frequ%
\^{e}ncia} $\delta $, que deve satisfazer a desigualdade $\left\vert \delta
\right\vert \ll \left\vert f_{1}\right\vert $, \'{e} obtido substituindo-se
a equa\c{c}\~ao (\ref{sol}) nas equa\c{c}\~{o}es (\ref{so1}) e (\ref{so2}).
Em ambos os casos, obtemos%
\begin{equation}
\delta ^{2}+f_{1}\delta -\sum_{i=1}^{\infty }p_{i}^{2}=0~,
\end{equation}%
e a solu\c{c}\~{a}o apropriada \'{e} 
\begin{equation}
\delta =\frac{\sqrt{f_{1}^{2}+4\sum_{i=1}^{\infty }p_{i}^{2}}-f_{1}}{2}%
\approx \frac{1}{f_{1}}\sum_{i=1}^{\infty }p_{i}^{2}\,.
\end{equation}

\subsubsection{Solu\c{c}\~{a}o anal\'{\i}tica final}

Nos casos de interesse, quando $L_{K}\neq 0$ para um n\'{u}mero inteiro $K$,
vamos supor que as solu\c{c}\~{o}es das equa\c{c}\~{o}es (\ref{apm}) sejam 
\begin{equation}
a_{+}=\alpha _{+}\left( t\right) e^{-it\delta }~,~a_{-}=\alpha _{-}\left(
t\right) e^{it\delta }\,,  \label{alf}
\end{equation}%
onde $\alpha _{\pm }\left( t\right) $ s\~{a}o fun\c{c}\~{o}es que variam
lentamente durante o intervalo de tempo $\tau $. Substituindo (\ref{alf})
nas equa\c{c}\~{o}es (\ref{apm}), e lembrando que as f\'ormulas (\ref{sol}) s%
\~{a}o as solu\c{c}\~{o}es aproximadas quando $L_{K}=0$, obtemos as
seguintes equa\c{c}\~{o}es diferenciais acopladas para as fun\c{c}\~{o}es $%
\alpha _{\pm }\left( t\right) $:%
\begin{eqnarray}
\dot{\alpha}_{+}\left( t\right) &=&-iL_{K}e^{iX_{K}t}\alpha _{-}\left(
t\right) \quad  \label{su3} \\
\dot{\alpha}_{-}\left( t\right) &=&-iL_{K}^{\ast }e^{-iX_{K}t}\alpha
_{+}\left( t\right) ~,  \label{su4}
\end{eqnarray}%
onde definimos 
\begin{equation}
X_{K}=R+2\delta -2\omega K\,.
\end{equation}

Estas equa\c{c}\~{o}es podem ser reescritas em forma de uma \'{u}nica equa%
\c{c}\~{a}o diferencial de segunda ordem com coeficientes constantes%
\begin{equation}
\ddot{\alpha}_{+}-iX_{K}\dot{\alpha}_{+}+\left\vert L_{K}\right\vert
^{2}\alpha _{+}=0~\,.  \label{su5}
\end{equation}%
Sua solu\c{c}\~{a}o \'{e}%
\begin{equation}
\alpha _{+}\left( t\right) =e^{iX_{K}t/2}\left[ S_{1}e^{i\Theta
_{K}t}+S_{2}e^{-i\Theta _{K}t}\right] ~,  \label{su}
\end{equation}%
onde%
\begin{equation}
\Theta _{K}=\sqrt{L_{K}^{2}+X^{2}/4}
\end{equation}%
e $S_{1}$ e $S_{2}$ s\~{a}o constantes determinadas pelas condi\c{c}\~{o}es
iniciais. Substituindo (\ref{su}) na equa\c{c}\~{a}o (\ref{su3}), encontramos%
\begin{equation}
\alpha _{-}\left( t\right) =-e^{-iX_{K}t/2}\left[ \frac{\left(
X_{K}/2+\Theta _{K}\right) }{L_{K}}S_{1}e^{i\Theta _{K}t}+\frac{\left(
X_{K}/2-\Theta _{K}\right) }{L_{K}}S_{2}e^{-i\Theta _{K}t}\right] \,.
\end{equation}%
Das condi\c{c}\~{o}es iniciais, encontramos%
\begin{eqnarray}
S_{1} &=&\frac{\left( \Theta _{K}-X_{K}/2\right) A_{+}\left( 0\right)
-L_{K}A_{-}\left( 0\right) }{2\Theta _{K}} \\
S_{2} &=&\frac{\left( \Theta _{K}+X_{K}/2\right) A_{+}\left( 0\right)
+L_{K}A_{-}\left( 0\right) }{2\Theta _{K}}\,.
\end{eqnarray}

Portanto, as solu\c{c}\~{o}es finais do nosso problema (perto das
resson\^ancias multifot\^onicas) s\~{a}o%
\begin{equation}
A_{+}(t)=e^{-i\Upsilon \sin \left( 2\omega t\right) /2}e^{i\left(
X_{K}/2-\delta \right) t}\left[ A_{+}\left( 0\right) \left( \cos \Theta
_{K}t-i\frac{X_{K}}{2\Theta _{K}}\sin \Theta _{K}t\right) -iA_{-}\left(
0\right) \frac{L_{K}}{\Theta _{K}}\sin \Theta _{K}t\right] \,~  \label{sc1}
\end{equation}%
\begin{equation}
A_{-}(t)=e^{i\Upsilon \sin \left( 2\omega t\right) /2}e^{-i\left(
X_{K}/2-\delta \right) t}\left[ A_{-}\left( 0\right) \left( \cos \Theta
_{K}t+i\frac{X_{K}}{2\Theta _{K}}\sin \Theta _{K}t\right) -iA_{+}\left(
0\right) \frac{L_{K}}{\Theta _{K}}\sin \Theta _{K}t\right] \,,~  \label{sc2}
\end{equation}%
e a probabilidade de encontrar o \'{a}tomo no estado excitado \'{e} dada
pela equa\c{c}\~{a}o (\ref{prob}). Para grandes dessintonias, $\Delta \gg G$%
, temos aproximadamente%
\begin{equation}
P_{e}(t)\approx \left\vert A_{+}(t)-\frac{G}{2\Delta }e^{iRt}A_{-}(t)\right%
\vert ^{2}~.  \label{gd}
\end{equation}

Para entender o que estas solu\c{c}\~{o}es representam fisicamente,
consideremos a condi\c{c}\~{a}o inicial $A_{-}\left( 0\right) =1$ e $%
A_{+}\left( 0\right) =0$, o que corresponde a $c_{g}\left( 0\right) \approx
1 $ e $c_{e}\left( 0\right) \approx 0$. Neste caso%
\begin{equation}
\left\vert A_{+}\left( t\right) \right\vert ^{2}=\frac{L_{K}^{2}}{\Theta
_{K}^{2}}\sin ^{2}\Theta _{K}t=\frac{L_{K}^{2}}{L_{K}^{2}+X_{K}^{2}/4}\sin
^{2}\Theta _{K}t\,
\end{equation}%
\begin{equation}
\left\vert A_{-}(t)\right\vert ^{2}=1-\frac{L_{K}^{2}}{L_{K}^{2}+X_{K}^{2}/4}%
\sin ^{2}\Theta _{K}t\,~.
\end{equation}%
Portanto, quando $X_{K}=0$, o \'{a}tomo tem 100\% de chance de passar do
estado fundamental para o estado excitado nos instantes do tempo iguais a m%
\'{u}ltiplos \'{\i}mpares de $T_{K}=\pi /(2\left\vert L_{K}\right\vert )$.
Isto corresponde justamente \`{a} resson\^{a}ncia de $\left( 2K+1\right) $-f%
\'{o}tons, pois este fen\^{o}meno ocorre quando a frequ\^{e}ncia de transi%
\c{c}\~{a}o at\^{o}mica \'{e} igual a%
\begin{equation}
\Omega _{K}=\omega +\sqrt{4\left( \omega K-\delta \right) ^{2}-G^{2}}\approx
\omega \left( 2K+1\right) -2\delta -\frac{G^{2}}{4\left( \omega K-\delta
\right) }~.
\end{equation}%
Se $X_{K}\neq 0$, a probabilidade de excita\c{c}\~{a}o do \'{a}tomo cai a
aproximadamente 50\% quando $X_{K}=\pm 2\left\vert L_{K}\right\vert $.
Portanto, podemos falar que a largura de linha total destas resson\^{a}ncias
multifot\^{o}nicas \'{e} igual a $4\left\vert L_{K}\right\vert $. Maiores
detalhes sobre esta e outras dedu\c{c}\~{o}es podem ser encontrados nas refer%
\^{e}ncias \cite{out,marinho2,out2}.

A tabela abaixo mostra os valores aproximados de $\Omega _{K}$, $L_{K}$ e $%
\omega T_{K}$ para $G=0.2\omega $:

\begin{center}
\begin{tabular}{|c|c|c|c|}
\hline
$\,K\,$ & $\,\Omega _{K}/\omega $\, & $\,\left\vert L_{K}\right\vert /\omega\, $ & \,$%
\omega T_{K}$\, \\ \hline
$1$ & $2.985$ & $2.5\times 10^{-4}$ & $6.3\times 10^{3}$ \\ \hline
$2$ & $4.992$ & $1.6\times 10^{-7}$ & $9.9\times 10^{6}$ \\ \hline
$3$ & $6.994$ & $4\times 10^{-11}$ & $4\times 10^{10}$ \\ \hline
$4$ & $8.995$ & $5\times 10^{-15}$ & $3\times 10^{14}$ \\ \hline
\end{tabular}%
\\[0pt]
\vspace{2mm} Tabela 1. Par\^ametros de resson\^ancias
multifot\^onicas para $G=0.2\omega $.
\end{center}

Vemos que as resson\^{a}ncias multifot\^{o}nicas s\~{a}o muito estreitas e
exigem tempos muito longos (da ordem de $T_{K}$) para serem observadas.
Logo, as taxas de dissipa\c{c}\~{a}o do \'{a}tomo e da cavidade devem ser
menores que $\left\vert L_{K}\right\vert $ para que as oscila\c{c}\~{o}es de 
$P_{e}$ possam ser observadas na pr\'{a}tica.

\begin{figure}[tbh]
\begin{center}
\includegraphics[width=0.99\textwidth]{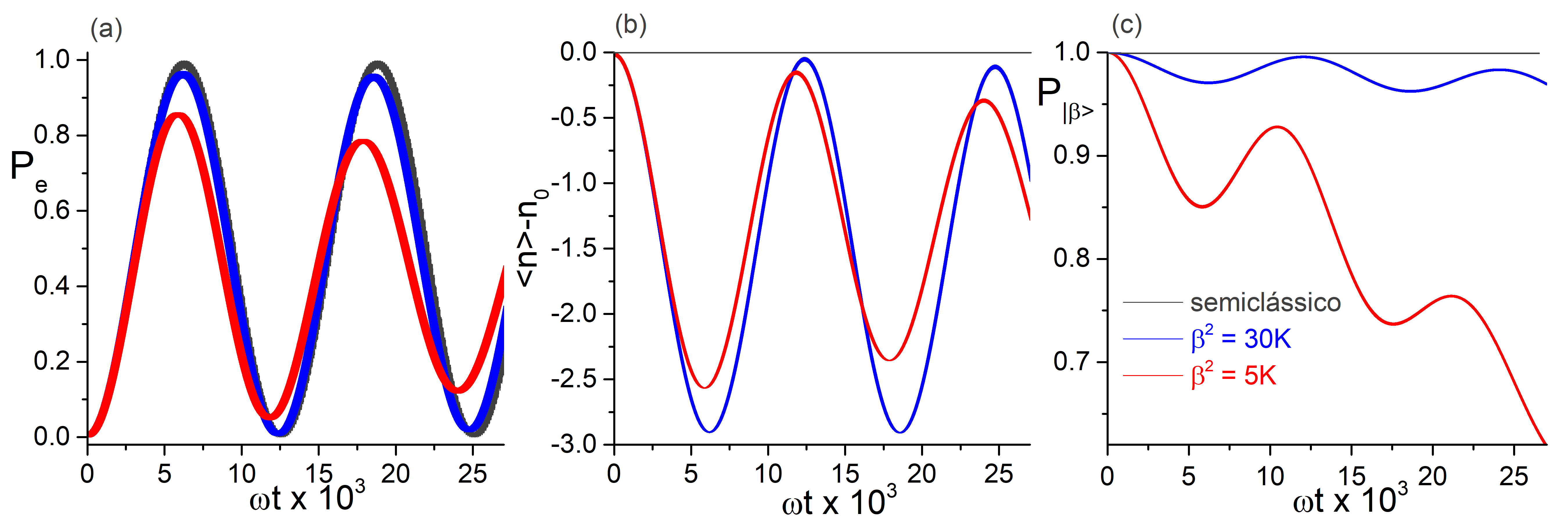} {}
\end{center}
\caption{Compara\c{c}\~{a}o de modelos de Rabi semicl\'{a}ssico e qu\^{a}%
ntico para $G=0.2\protect\omega $ e $\Omega =2.98497\protect\omega $, quando
a excita\c{c}\~{a}o do \'{a}tomo acontece via absor\c{c}\~{a}o de tr\^{e}s f%
\'{o}tons. a) Probabilidade de excita\c{c}\~{a}o do \'{a}tomo, partindo do
estado inicial $|g\rangle $, para o modelo semicl\'{a}ssico (linha cinza) e
modelo qu\^{a}ntico com $\protect\beta ^{2}=5$K (linha vermelha) e $\protect%
\beta ^{2}=30$K (linha azul), onde $\protect\beta $ \'{e} a amplitude do estado
coerente $|\mathbf{\protect\beta }\rangle $. b) Varia\c{c}\~{a}o do n\'{u}%
mero m\'{e}dio de f\'{o}tons na cavidade. c) Probabilidade de que o campo
eletromagn\'{e}tico permanece no estado coerente inicial $|\mathbf{\protect%
\beta }\rangle $. }
\label{fig4}
\end{figure}

O comportamento de $P_{e}$ no regime de resson\^{a}ncia de tr\^{e}s f\^{o}%
tons est\'{a} ilustrado na figura \ref{fig4}a, em que a linha cinza mostra a
solu\c{c}\~{a}o (\ref{gd}) em fun\c{c}\~{a}o do tempo para os par\^{a}metros 
$G=0.2\omega $ e $\Omega =2.98497\omega $. Esta curva \'{e} praticamente
indistingu\'{\i}vel da solu\c{c}\~{a}o num\'{e}rica exata das equa\c{c}\~{o}%
es (\ref{f1}) -- (\ref{f2}) (n\~{a}o mostradas), o que atesta a precis\~{a}o
do nosso m\'{e}todo anal\'{\i}tico. Da Tabela 1 encontramos o intervalo de
tempo necess\'ario para a excita\c{c}\~{a}o competa do \'{a}tomo, $\omega T_{1}=\omega
\pi /2\left\vert L_{1}\right\vert \approx 6.3$K (aqui K denota $10^{3}$), o
que est\'{a} em perfeito acordo com os dados num\'{e}ricos. Notem como este
processo \'{e} lento comparado \`{a} excita\c{c}\~{a}o do \'{a}tomo via
processo de um f\'{o}ton (quando $\Delta =0$), para o qual ter\'{\i}amos $%
\omega T_{\pi }=\pi \omega /G\approx 16$. No entanto, a pr\'{o}pria
possibilidade de excitar o \'{a}tomo para $\Omega \approx 3\omega $, mesmo
no regime semicl\'{a}ssico, \'{e} bastante interessante e, na nossa opini%
\~{a}o, merece ser mencionada quando se estuda a intera\c{c}\~{a}o radia\c{c}%
\~{a}o-mat\'{e}ria!

\section{Modelo de Rabi Qu\^{a}ntico\label{QRMz}}

O modelo de Rabi qu\^{a}ntico \cite{rev,solano,braak} \'{e} dado pelo
Hamiltoniano%
\begin{equation}
\hat{H}_{Q}=\frac{\hbar \Omega }{2}\hat{\sigma}_{z}+\hbar \omega \hat{n}%
+\hbar g\left( \hat{a}+\hat{a}^{\dagger }\right) \left( \hat{\sigma}_{+}+%
\hat{\sigma}_{-}\right) \,,  \label{QRM}
\end{equation}%
onde $\hat{a}$ e $\hat{a}^{\dagger }$ s\~{a}o os operadores de aniquila\c{c}%
\~{a}o e cria\c{c}\~{a}o do modo de campo eletromagn\'{e}tico com a frequ%
\^{e}ncia $\omega $, respectivamente. $\hat{n}=\hat{a}^{\dagger }\hat{a}$ 
\'{e} o operador-n\'{u}mero, cujo valor m\'{e}dio fornece o n\'{u}mero m\'{e}%
dio de f\'{o}tons na cavidade. $g$ \'{e} a constante de acoplamento entre o 
\'{a}tomo e o campo eletromagn\'{e}tico quantizado, sendo diferente da
constante de acoplamento $G$ no modelo semicl\'{a}ssico. O modelo de Rabi qu%
\^{a}ntico foi resolvido analiticamente em 2011 por Daniel Braak, \cite%
{braak}, mas a solu\c{c}\~{a}o exata exige encontrar zeros de uma fun\c{c}%
\~{a}o transcendente, definida por uma s\'{e}rie infinita. Antes disso,
diversos m\'{e}todos de solu\c{c}\~{o}es aproximadas haviam sido desenvolvidos. Uma das aproxima\c{c}\~{o}es mais usadas consiste na omiss%
\~{a}o dos chamados \textquotedblleft termos
contragirantes\textquotedblright , $\left( \hat{a}\hat{\sigma}_{-}+\hat{a}%
^{\dagger }\hat{\sigma}_{+}\right) $, o que resulta no modelo de
Jaynes-Cummings \cite{JC}, proposto em 1963. A solu\c{c}\~{a}o exata do
modelo de Jaynes-Cummings \'{e} bem simples e consegue explicar uma grande
variedade de fen\^{o}menos envolvendo a intera\c{c}\~{a}o entre a luz e a mat%
\'{e}ria quando a constante de acoplamento \'{e} muito pequena comparada a $%
\omega $ e $\Omega $ \cite{scully,schleich,knight,orszag,larson}. Neste
trabalho, por\'{e}m, vamos resolver exatamente o Hamiltoniano (\ref{QRM})
usando m\'{e}todos num\'{e}ricos.

Para ver como o Hamiltoniano de Rabi qu\^{a}ntico se relaciona com o semicl%
\'{a}ssico, equa\c{c}\~{a}o (\ref{sr}), vamos definir um novo estado do
sistema: $|\varphi (t)\rangle =\exp \left( i\omega \hat{n}t\right) |\psi
(t)\rangle $ [onde $|\psi (t)\rangle $ obedece \`{a} equa\c{c}\~{a}o de Schr%
\"{o}dinger com o Hamiltoniano (\ref{QRM})]. Com isso, encontramos que $%
|\varphi (t)\rangle $ obedece \`{a} equa\c{c}\~{a}o de Schr\"{o}dinger com o
Hamiltoniano (\ref{qr}), ou seja, o Hamiltoniano (\ref{qr}) \'{e}
simplesmente o Hamiltoniano (\ref{QRM}) escrito num quadro de
intera\c{c}\~{a}o diferente. O conjunto dos estados de Fock $|n\rangle $, onde $%
n=0,1,2,\ldots $, forma uma base ortonormal de dimens\~{a}o infinita que
gera o espa\c{c}o de Hilbert correspondente ao modo do campo eletromagn\'{e}%
tico na cavidade. Ent\~{a}o, podemos expandir o estado $|\varphi (t)\rangle $
como%
\begin{equation}
|\varphi (t)\rangle =\sum_{n=0}^{\infty }\left[ A_{n}(t)|g\rangle \otimes
|n\rangle \,+B_{n}(t)|e\rangle \otimes |n\rangle \,\right] ~,  \label{state}
\end{equation}%
onde $\otimes $ denota o produto tensorial, e $A_{n}$ e $B_{n}$ s\~{a}o as
amplitudes de probabilidade de estados $|g\rangle \otimes |n\rangle $ e $%
|e\rangle \otimes |n\rangle $ do sistema bipartido \'{a}tomo-campo.
Suponhamos que, no instante inicial, o \'{a}tomo estava no estado $|\varphi
_{at}\rangle $ e o campo eletromagn\'{e}tico foi preparado no \emph{estado
coerente }\cite{scully,schleich} 
\begin{equation}
|\mathbf{\beta }\rangle =e^{-\beta ^{2}/2}\sum_{n=0}^{\infty }\frac{\beta
^{n}}{\sqrt{n!}}|n\rangle \,.
\end{equation}%
Aqui, para diferenciar o estado $|\mathbf{\beta }\rangle $ de sua amplitude $%
\beta $, usamos o negrito para denotar o estado coerente; ademais, sem perda
de generalidade, supomos que o par\^{a}metro $\beta $ \'{e} real. O estado
coerente possui as seguintes propriedades importantes: 
\begin{equation}
\hat{a}|\mathbf{\beta }\rangle =\beta |\mathbf{\beta }\rangle ~,~\langle 
\mathbf{\beta }|\hat{a}^{\dagger }=\beta ^{\ast }\langle \mathbf{\beta }%
|~,~\langle \mathbf{\beta }|\mathbf{\beta }\rangle =1~.
\end{equation}

Vejamos qual ser\'{a} o estado do sistema ap\'os um tempo infinitesimal $dt$, para
o estado inicial $|\varphi \left( 0\right) \rangle =|\varphi _{at}\rangle
\otimes |\mathbf{\beta }\rangle $. Da Equa\c{c}\~{a}o de Schr\"{o}dinger com
o Hamiltoniano $\hat{H}^{\prime }$, equa\c{c}\~{a}o (\ref{qr}), obtemos%
\begin{equation}
|\varphi \left( dt\right) \rangle =|\varphi \left( 0\right) \rangle -\frac{i%
\hat{H}^{\prime }dt}{\hbar }|\varphi \left( 0\right) \rangle ~.
\end{equation}%
O valor m\'{e}dio de qualquer observ\'{a}vel do \'{a}tomo, $\hat{O}_{at}$,
no tempo $dt$ \'{e}%
\begin{eqnarray}
\left\langle \hat{O}_{at}\right\rangle \left( dt\right) &=&\langle \varphi
\left( dt\right) |\hat{O}_{at}\otimes \mathbf{I}_{cam}|\varphi \left(
dt\right) \rangle \\
&=&\langle \varphi \left( 0\right) |\hat{O}_{at}\otimes \mathbf{I}%
_{cam}|\varphi \left( 0\right) \rangle -idt\langle \varphi \left( 0\right) |%
\hat{O}_{at}\otimes \mathbf{I}_{cam}\frac{\hat{H}^{\prime }}{\hbar }|\varphi
\left( 0\right) \rangle +idt\langle \varphi \left( 0\right) |\frac{\hat{H}%
^{\prime }}{\hbar }\hat{O}_{at}\otimes \mathbf{I}_{cam}|\varphi \left(
0\right) \rangle ~,  \notag
\end{eqnarray}%
onde $\mathbf{I}_{cam}$ \'{e} o operador-identidade no subespa\c{c}o de
Hilbert do campo. No nosso caso, obtemos%
\begin{equation}
\langle \varphi \left( 0\right) |\hat{O}_{at}|\varphi \left( 0\right)
\rangle =\langle \mathbf{\beta }|\otimes \langle \varphi _{at}|\hat{O}%
_{at}\otimes \mathbf{I}_{cam}|\varphi _{at}\rangle \otimes |\mathbf{\beta }%
\rangle =\langle \varphi _{at}|\hat{O}_{at}|\varphi _{at}\rangle
\end{equation}%
\begin{eqnarray}
\langle \varphi \left( 0\right) |\hat{O}_{at}\otimes \mathbf{I}_{cam}\frac{%
\hat{H}^{\prime }}{\hbar }|\varphi \left( 0\right) \rangle &=&\langle 
\mathbf{\beta }|\otimes \langle \varphi _{at}|\hat{O}_{at}\otimes \mathbf{I}%
_{cam}\left[ \frac{\Omega }{2}\hat{\sigma}_{z}+g\left( \hat{a}e^{-i\omega t}+%
\hat{a}^{\dagger }e^{i\omega t}\right) \left( \hat{\sigma}_{+}+\hat{\sigma}%
_{-}\right) \right] |\varphi _{at}\rangle \otimes |\mathbf{\beta }\rangle 
\notag \\
&=&\langle \varphi _{at}|\hat{O}_{at}\left[ \frac{\Omega }{2}\hat{\sigma}%
_{z}+g\left( \beta e^{-i\omega t}+\beta ^{\ast }e^{i\omega t}\right) \left( 
\hat{\sigma}_{+}+\hat{\sigma}_{-}\right) \right] |\varphi _{at}\rangle
\end{eqnarray}%
e um termo an\'{a}logo para $\langle \varphi \left( 0\right) |\frac{\hat{H}%
^{\prime }}{\hbar }\hat{O}\otimes \mathbf{I}_{cam}|\varphi \left( 0\right)
\rangle $. Com isso, vemos que para tempos iniciais, a din\^{a}mica do \'{a}%
tomo \'{e} regida pelo Hamiltoniano semicl\'{a}ssico (\ref{sr}) com a
constante de acoplamento%
\begin{equation}
G=2g\beta ~.  \label{ggg}
\end{equation}%
Uma abordagem mais rigorosa, analisando a transi\c{c}\~{a}o do Hamiltoniano
de Rabi qu\^{a}ntico para o semicl\'{a}ssico para campos coerentes intensos,
foi dada nos artigos recentes \cite{shus1,shus2}.

O nosso objetivo aqui ser\'{a} resolver numericamente a Equa\c{c}\~{a}o de
Schr\"{o}dinger para o Hamiltoniano (\ref{QRM}) e comparar os resultados
semicl\'{a}ssicos com os qu\^{a}nticos, empregando a rela\c{c}\~{a}o (\ref%
{ggg}), que relaciona as constantes de acoplamento semicl\'{a}ssica e qu\^{a}%
ntica. Para isso, substituindo as equações (\ref{qr}) e (\ref{state}) na equação de Schrödinger,
\begin{equation}
i\hbar\frac{\partial}{\partial t}\left|\varphi\left(t\right)\right\rangle =\hat{H}^{\prime}\left|\varphi\left(t\right)\right\rangle,
\label{ben}
\end{equation}
encontramos as equações diferenciais acopladas que precisamos resolver numericamente:
\begin{equation}
i\dot{A}_{m}=-\frac{\Omega }{2}A_{m}+g\left( e^{-i\omega t}\sqrt{m+1}%
B_{m+1}+e^{i\omega t}\sqrt{m}B_{m-1}\right)  \label{ta1}
\end{equation}%
\begin{equation}
i\dot{B}_{m}\,=\frac{\Omega }{2}B_{m}+g\left( e^{-i\omega t}\sqrt{m+1}%
A_{m+1}+e^{i\omega t}\sqrt{m}A_{m-1}\right)\,,  \label{ta2}
\end{equation}%
[para obter (\ref{ta1}) atuamos $\left\langle g,m\right|$, e para obter (\ref{ta2}) atuamos $\left\langle e,m\right|$ do lado esquerdo da Eq. (\ref{ben})] com as condi\c{c}\~{o}es iniciais $B_{m}(0)=0$ e $A_{m}(0)=e^{-\beta
^{2}/2}\beta ^{m}/\sqrt{m!}$, onde $m$ varia de $0$ a $\infty $. Por\'{e}m,
podemos simplificar o problema impondo que $A_{m}\left( t\right) =0$ para $%
m<N_{1}$ e $m>N_{2}$, onde $N_{1}$ e $N_{2}$ s\~{a}o n\'{u}meros inteiros.
Isto decorre do fato que, para $\beta \gg 1$ (o que corresponde a um estado
com muitos f\'{o}tons -- isto \'e, um estado aproximadamente
\textquotedblleft cl\'{a}ssico\textquotedblright ), a probabilidade de
encontrar $k$ f\'{o}tons no estado coerente $|\mathbf{\beta }\rangle $ \'{e}
dada por uma distribui\c{c}\~{a}o Gaussiana \cite{scully}%
\begin{equation}
\left\vert \langle k|\beta \rangle \right\vert ^{2}\simeq \frac{1}{\sqrt{%
2\pi }\Delta n}\exp \left[ -\left( \frac{k-\left( \left\langle \hat{n}%
\right\rangle -1/2\right) }{\sqrt{2}\Delta n}\right) ^{2}\right] \,,
\label{di}
\end{equation}%
onde $\left\langle \hat{n}\right\rangle =\langle \mathbf{\beta }|\hat{a}%
^{\dagger }\hat{a}|\mathbf{\beta }\rangle =\beta ^{2}$ \'{e} o n\'{u}mero m%
\'{e}dio de f\'{o}tons e $\Delta n=\sqrt{\left\langle \hat{n}%
^{2}\right\rangle -\left\langle \hat{n}\right\rangle ^{2}}$ \'{e} o desvio
padr\~{a}o. Para o estado coerente, usando a rela\c{c}\~{a}o de comuta\c{c}%
\~{a}o entre $\hat{a}$ e $\hat{a}^{\dagger }$, obtemos%
\begin{equation}
\left\langle \hat{n}^{2}\right\rangle =\langle \mathbf{\beta }|\hat{a}%
^{\dagger }\hat{a}\hat{a}^{\dagger }\hat{a}|\mathbf{\beta }\rangle =\langle 
\mathbf{\beta }|\hat{a}^{\dagger }\left( \hat{a}^{\dagger }\hat{a}+1\right) 
\hat{a}|\mathbf{\beta }\rangle =\beta ^{4}+\beta ^{2}~.
\end{equation}%
Com isso, obtemos $\Delta n=\sqrt{\left\langle \hat{n}^{2}\right\rangle
-\left\langle \hat{n}\right\rangle ^{2}}=\sqrt{\beta ^{4}+\beta ^{2}-\beta
^{4}}=\beta $. Vemos que a distribui\c{c}\~{a}o (\ref{di}) \'{e} centrada em 
$\beta ^{2}-1/2$ e tende a zero para valores de $k$ tais que $\left\vert
k-\beta ^{2}\right\vert \gg \sqrt{2}\beta $. Como neste trabalho n\~{a}o h%
\'{a} cria\c{c}\~{a}o nem destrui\c{c}\~{a}o significativa de f\'{o}tons,
podemos supor, para fins num\'{e}ricos, que os estados de Fock com $n<N_{1}$
e $n>N_{2}$ possuem probabilidade t\~{a}o baixa (menor que $10^{-30}$,
digamos), que podemos desprez\'{a}-los completamente! Por exemplo, para $%
\beta ^{2}=5\times 10^{3}$ escolhemos $N_{1}=4145$, $N_{2}=5906$, e para $%
\beta ^{2}=3\times 10^{4}$ escolhemos $N_{1}=27879$, $N_{2}=32171$.

Ao resolver numericamente as equa\c{c}\~{o}es (\ref{ta1}) -- (\ref{ta2}),
podemos calcular imediatamente: a probabilidade do estado excitado do \'{a}%
tomo%
\begin{equation}
P_{e}=\sum_{n=N_{1}}^{N_{2}}\left\vert \langle e,n|\varphi \left( t\right)
\rangle \right\vert ^{2}=\sum_{n=N_{1}}^{N_{2}}\left\vert B_{n}\left(
t\right) \right\vert ^{2}\,~,  \label{difr}
\end{equation}%
o n\'{u}mero m\'{e}dio de f\'{o}tons%
\begin{equation}
\left\langle \hat{n}\right\rangle =\sum_{n=N_{1}}^{N_{2}}n\left( \left\vert
\langle e,n|\varphi \left( t\right) \rangle \right\vert ^{2}+\left\vert
\langle g,n|\varphi \left( t\right) \rangle \right\vert ^{2}\right)
=\sum_{n=N_{1}}^{N_{2}}n\left( \left\vert A_{n}\left( t\right) \right\vert
^{2}+\left\vert B_{n}\left( t\right) \right\vert ^{2}\right)
\end{equation}%
e a probabilidade de que o campo eletromagn\'{e}tico permanece no estado
inicial $|\beta \rangle $ (independentemente de o \'{a}tomo estar no estado
excitado ou fundamental), que chamamos de $P_{|\beta \rangle }$%
\begin{equation}
P_{|\beta \rangle }=\left\vert \langle g,\mathbf{\beta }|\varphi (t)\rangle
\right\vert ^{2}+\left\vert \langle e,\mathbf{\beta }|\varphi (t)\rangle
\right\vert ^{2}=e^{-\beta ^{2}}\left[ \left\vert \sum_{n=0}^{\infty }\frac{%
\beta ^{n}}{\sqrt{n!}}A_{n}(t)\right\vert ^{2}+\left\vert \sum_{n=0}^{\infty
}\frac{\beta ^{n}}{\sqrt{n!}}B_{n}(t)\right\vert ^{2}\right] \,.
\end{equation}%
Com isto, poderemos caracterizar a din\^{a}mica do sistema \'{a}tomo-campo
no regime qu\^{a}ntico.

\subsection{Comportamento do sistema no formalismo qu\^{a}ntico}

N\'{o}s resolvemos numericamente as equa\c{c}\~{o}es diferenciais acopladas (%
\ref{ta1}) e (\ref{ta2}) usando o m\'{e}todo de Runge-Kutta-Verner de quinta
e sexta ordens, e comparamos estas solu\c{c}\~{o}es com as solu\c{c}\~{o}es
semicl\'{a}ssicas (\ref{sc1}) e (\ref{sc2}) [que coincidiram com as solu\c{c}%
\~{o}es num\'{e}ricas exatas do Hamiltoniano (\ref{sr}) para os par\^{a}%
metros escolhidos]. Na figura \ref{fig3} escolhemos o estado inicial $%
|g\rangle \otimes |\beta \rangle $ com o n\'{u}mero m\'{e}dio de f\'{o}tons $%
\beta ^{2}=5$K (linhas vermelhas) e $\beta ^{2}=30$K (linhas azuis); os
demais par\^{a}metros foram $\omega =\Omega $ e $g=10^{-2}\pi /\beta $, o
que corresponde \`{a} dura\c{c}\~{a}o do pulso-$\pi $ de $\omega T_{\pi }=50$%
. Analisando $P_{e}$ na figura \ref{fig3}a, vemos que quanto maior o valor
de $\beta $, mais a din\^{a}mica semicl\'{a}ssica (linha cinza) se aproxima da din\^{a}mica
qu\^{a}ntica (ou seja, a correta) para tempos iniciais: por exemplo, para $\beta ^{2}=30$ K,
as primeiras cinco oscila\c{c}\~{o}es de Rabi s\~{a}o praticamente id\^{e}%
nticas nos dois modelos, enquanto
para $\beta ^{2}=5$K apenas as duas primeiras oscila\c{c}\~{o}es coincidem. No entanto, para tempos
maiores, o formalismo qu\^{a}ntico leva inevitavelmente ao colapso de oscila%
\c{c}\~{o}es de $P_{e}$ devido \`{a} interfer\^{e}ncia destrutiva entre as
diferentes probabilidades $\left\vert B_{n}\right\vert ^{2}$ na equa\c{c}%
\~{a}o (\ref{difr}) -- este \'{e} o famoso fen\^{o}meno de {\em colapso e
ressurgimento} da invers\~{a}o at\^{o}mica \cite{eberly,scully,acosta1}, que n%
\~{a}o ocorre no formalismo semicl\'{a}ssico.

A figura \ref{fig3}b mostra a varia\c{c}\~{a}o do n\'{u}mero m\'{e}dio de f%
\'{o}tons na cavidade de acordo com o modelo qu\^{a}ntico. Como esperado, a
excita\c{c}\~{a}o do \'{a}tomo \'{e} acompanhada pela absor\c{c}\~{a}o de um
f\'{o}ton do campo eletromagn\'{e}tico, que \'{e} reabsorvido pelo campo
quando o \'{a}tomo \'{e} desexcitado; o n\'{u}mero m\'{e}dio de f\'{o}tons
tamb\'{e}m exibe o fen\^{o}meno de colapso, pois a probabilidade
de excita\c{c}\~{a}o at\^{o}mica deixa de oscilar com o passar do tempo. A
linha cinza indica a aus\^encia de varia\c{c}\~{a}o no n\'{u}mero m\'{e}dio
de f\'{o}tons (proporcional \`a energia do campo eletromagn%
\'{e}tico) no regime semicl\'{a}ssico, j\'{a} que este negligencia qualquer modifica\c{c}\~{a}o do campo
devido \`{a} intera\c{c}\~{a}o com o \'{a}tomo. Na figura \ref{fig3}c
mostramos a probabilidade de encontrar o campo no estado inicial $|\beta
\rangle $ (que seria igual a um no modelo semicl\'{a}ssico, como indicado
pela linha cinza). No modelo qu\^{a}ntico, esta probabilidade diminui com
tempo, a diminui\c{c}\~{a}o sendo mais r\'{a}pida para o n\'{u}mero de f\'{o}%
tons menor (embora não mostrado aqui, esta probabilidade pode vir a aumentar esporadicamente para tempos muito mais longos, devido ao fen\^omeno de ressurgimento). Isto comprova, mais uma vez, que o estado do campo eletromagn%
\'{e}tico \'{e} severamente modificado depois de algumas oscila\c{c}\~{o}es
de Rabi; mesmo assim, para as primeiras poucas oscila\c{c}\~{o}es de Rabi, a altera%
\c{c}\~{a}o no estado do campo \'{e} pequena e pode ser desprezada para
grandes valores de $\beta ^{2}$. Maiores detalhes sobre  modifica\c{c}\~{o}%
es na distribui\c{c}\~ao do n\'umero de f\'{o}tons e o grau de emaranhamento
entre o \'{a}tomo e o campo podem ser encontrados na refer\^{e}ncia \cite%
{acosta1}.

Na figura \ref{fig4} fazemos uma an\'{a}lise similar para a resson\^{a}ncia
de tr\^{e}s f\'{o}tons, considerando os par\^{a}metros $\Omega
=2.98497\omega $ e $g=0.1\omega /\beta $ (o que corresponde \`{a} constante
de acoplamento semicl\'{a}ssico $G=0.2\omega $). Na figura \ref{fig4}a,
vemos que para $\beta ^{2}=30$K o comportamento de $P_{e}$ no regime qu\^{a}%
ntico quase coincide com o comportamento semicl\'{a}ssico; por\'{e}m, para $%
\beta ^{2}=5$K a concord\^{a}ncia \'{e} bem pior, e o uso do modelo semicl%
\'{a}ssico neste caso n\~{a}o \'{e} apropriado. A figura \ref{fig4}b mostra
a varia\c{c}\~{a}o no n\'{u}mero m\'{e}dio de f\'{o}tons, comprovando que a
excita\c{c}\~{a}o at\^{o}mica acontece \`{a}s custas de absor\c{c}\~{a}o de
tr\^{e}s f\'{o}tons da cavidade. O n\'{u}mero m\'{e}dio de f\'{o}tons
absorvidos \'{e} menor que tr\^{e}s, e a probabilidade de excita\c{c}\~{a}o
at\^{o}mica \'{e} menor que um. Isto ocorre porque, mesmo para $\beta
^{2}=30 $K, nem todas as amplitudes de probabilidade $A_{n}$ se acoplam
de forma ressonante a $B_{m}$ (para valores de $n$ variando entre $\beta
^{2}-\beta $ e $\beta ^{2}+\beta $), de modo que a soma na equa\c{c}\~{a}o (%
\ref{difr}) n\~{a}o consegue alcan\c{c}ar o valor m\'{a}ximo de um. Al\'{e}m
disso, notamos que ambos $P_{e}$ e $\left\langle \hat{n}\right\rangle $
exibem o fen\^{o}meno de colapso nas oscila\c{c}\~{o}es devido \`{a} interfer%
\^{e}ncia destrutiva, e a probabilidade de perman\^{e}ncia do campo no
estado inicial $|\beta \rangle $ diminui com o passar do tempo.

\begin{figure}[tbh]
\begin{center}
\includegraphics[width=0.59\textwidth]{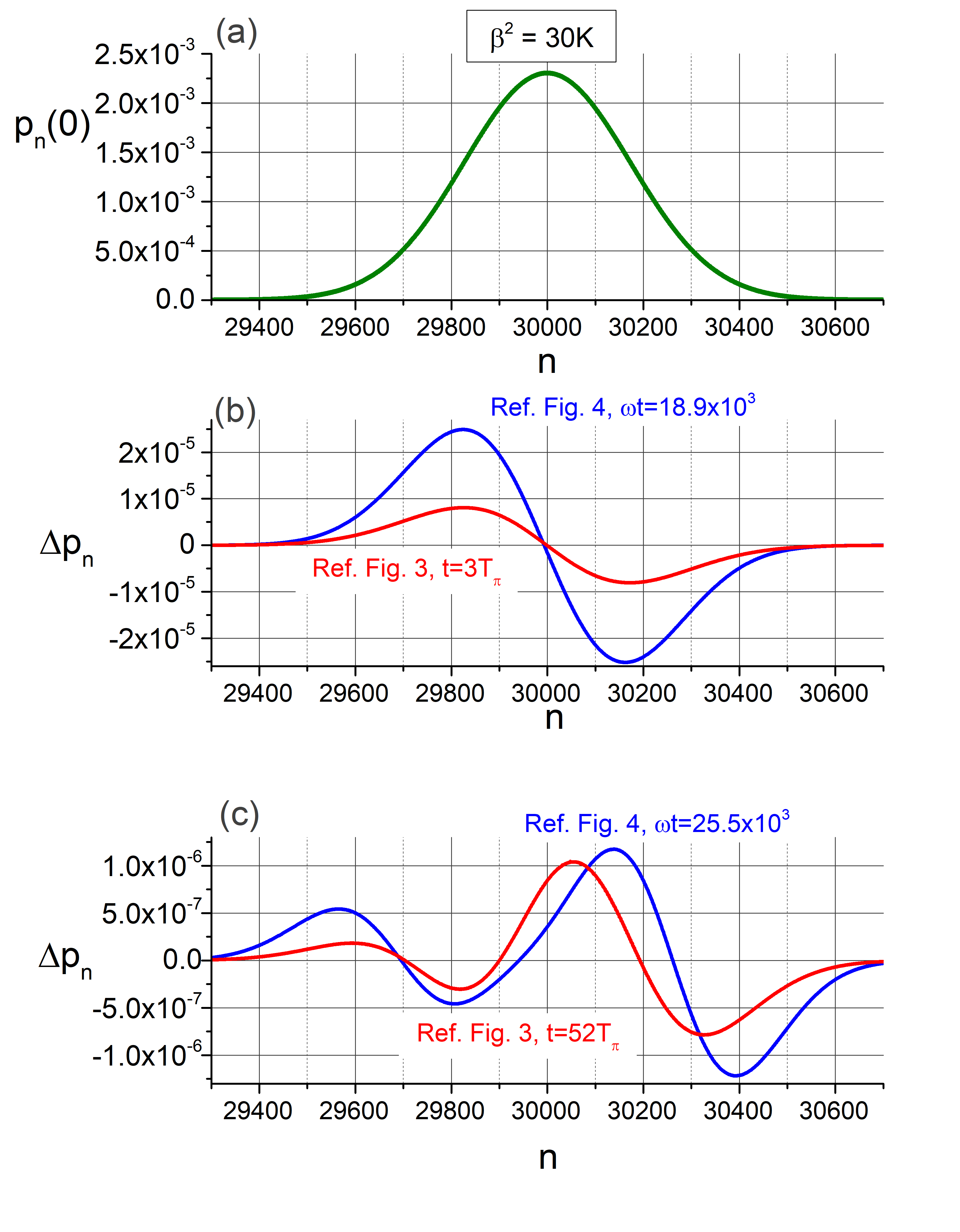} {}
\end{center}
\caption{a) Distribuição de probabilidade do número de fótons no estado inicial coerente $|\beta\rangle$, com $\beta^2=30$K. b) Diferença entre a distribuição de probabilidade de fótons nos instantes de tempo $t$ e a distribuição de fótons inicial, $\Delta p_n$, quando a probabilidade de excitação do átomo é próxima de $100\%$. c) $\Delta p_n$ quando a probabilidade de excitação do átomo é próxima de $0\%$. Curvas vermelhas (azuis) correspondem aos parâmetros da Figura \ref{fig3} (Figura \ref{fig4}).}.
\label{fig5}
\end{figure}

Estas duas figuras comprovam que quanto maior for o n\'{u}mero m\'{e}dio de
f\'{o}tons na cavidade, mais precisa fica a descri\c{c}\~{a}o semicl\'{a}ssica
para tempos iniciais, pois a probabilidade de perman\^{e}ncia do campo no
estado inicial permanece pr\'{o}xima de 100\%, e a varia\c{c}\~{a}o no n\'{u}%
mero m\'{e}dio de f\'{o}tons \'{e} muito pequena em rela\c{c}\~{a}o ao valor
inicial. No entanto, para tempos grandes (a escala de tempo depende do n\'{u}mero m%
\'{e}dio de f\'{o}tons na cavidade), o modelo semicl\'{a}ssico,
inevitavelmente, entra em desacordo com o modelo qu\^{a}ntico.

Por último, na Figura \ref{fig5} mostramos de que forma, especificamente, a interação com o átomo altera a distribuição de probabilidade do número de fótons do campo na cavidade. A Figura \ref{fig5}a mostra as probabilidades, $p_n(0)$, de haver $n$ fótons no estado inicial coerente do campo, $|\beta\rangle$, em que supomos que $\beta^2=30\times 10^3$. A Figura \ref{fig5}b mostra a diferença entre a distribuição de probabilidade no tempo $t$ e no tempo inicial, $\Delta p_n=p_n(t)-p_n(0)$, nos instantes em que o átomo tem a probabilidade de quase $100\%$ de estar no estado excitado. A linha vermelha corresponde aos parâmetros da Figura \ref{fig3} (ressonância de 1-fóton) e o instante de tempo $t=3T_\pi$; a linha azul corresponde aos parâmetros da Figura \ref{fig4} (ressonância de 3-fótons) e o instante de tempo $t=18.9\times 10^3 \omega^{-1}$. Vemos que, em ambos os casos, a excitação do átomo ocorre às custas de uma ligeira diminuição na probabilidade de haver mais que $\beta^2$ fótons na cavidade, e um respectivo aumento na probabilidade de haver menos que $\beta^2$ fótons. Ademais, a variação de $\Delta p_n$ é maior para a resonância de 3-fótons, embora não passe de $3\times 10^{-5}$ neste exemplo. A Figura \ref{fig5}c mostra $\Delta p_n$ nos instantes de tempo em que o átomo está aproximadamente no estado fundamental, ou seja, quando os fótons previamente absorvidos pelo átomo foram reemitidos para a cavidade. A linha vermelha corresponde aos parâmetros da Figura \ref{fig3} e o instante do tempo $t=52T_\pi$; a linha azul corresponde aos parâmetros da Figura \ref{fig4} e o instante do tempo $t=25.5\times10^3\omega^{-1}$. Vemos que $\Delta p_n$ é não nulo (embora não passe de $10^{-6}$ neste exemplo), ou seja, mesmo quando o átomo volta para o estado inicial $|g\rangle$, o estado do campo eletromagnético não retorna a seu estado inicial (em total acordo com as Figuras \ref{fig3}c e \ref{fig4}c). No entanto, se a variação $\Delta p_n$ puder ser desprezada para todos os instantes do tempo de interesse (como acontece para $\beta^2\gg 1$ e tempos relativamente curtos), recuperamos (aproximadamente) o modelo semiclássico.

\section{Conclus\~{o}es\label{conc}}

Neste artigo, realizamos uma dedu\c{c}\~{a}o concisa dos modelos de Rabi
semicl\'{a}ssico e qu\^{a}ntico, que descrevem a intera\c{c}\~{a}o entre um 
\'{a}tomo de dois n\'{\i}veis e o campo eletromagn\'{e}tico, tratado como uma fun\c{c}\~{a}o cl\'{a}ssica ou como um campo bos%
\^{o}nico quantizado, respectivamente. Apresentamos a solu\c{c}\~{a}o anal\'{\i}tica da evolu%
\c{c}\~{a}o temporal do estado qu\^{a}ntico do \'{a}tomo no modelo semicl%
\'{a}ssico e discutimos como as resson\^{a}ncias de um e m\'{u}ltiplos f\'{o}%
tons surgem naturalmente nesse contexto. Em seguida, mostramos como estudar
numericamente o modelo de Rabi qu\^{a}ntico no regime semicl\'{a}ssico,
quando o campo eletromagn\'{e}tico se encontra inicialmente em um estado
coerente com alto n\'{u}mero m\'{e}dio de f\'{o}tons. Verificamos que, para
tempos curtos, os resultados semicl\'{a}ssicos concordam com os qu\^{a}%
nticos, desde que o n\'{u}mero m\'{e}dio de f\'{o}tons seja suficientemente
alto, e excita\c{c}\~{a}o e desexcita\c{c}\~{a}o peri\'{o}dicas do \'{a}tomo
s\~{a}o acompanhados de absor\c{c}\~{a}o e emiss\~{a}o de um n\'{u}mero 
\'{\i}mpar de f\'{o}tons. Contudo, para tempos mais longos, o modelo semicl%
\'{a}ssico deixa de reproduzir o comportamento do modelo qu\^{a}ntico, no
qual o n\'{u}mero m\'{e}dio de f\'{o}tons e a probabilidade de excita\c{c}%
\~{a}o do \'{a}tomo deixam de exibir oscila\c{c}\~{o}es devido \`{a} interfer%
\^{e}ncia qu\^{a}ntica destrutiva. Al\'{e}m disso, o estado do campo
eletromagn\'{e}tico sofre modifica\c{c}\~{o}es significativas, de modo que,
para tempos longos, a probabilidade de ele permanecer no estado inicial
fica pr\'oxima de zero.

O formalismo descrito neste artigo \'{e} bastante intuitivo, envolvendo
apenas a solu\c{c}\~{a}o aproximada de um sistema de duas equa\c{c}\~{o}es
diferenciais ordin\'{a}rias por meio de algumas transforma\c{c}\~{o}es
simples, e pode ser estendido para descrever outros fen\^{o}menos de intera%
\c{c}\~{a}o entre radia\c{c}\~{a}o e mat\'{e}ria.

\section*{Agradecimentos}

Alexandre P. Costa agradece \`{a} Coordena\c{c}\~{a}o de Aperfei\c{c}oamento
de Pessoal de N\'{\i}vel Superior pelo apoio financeiro (C\'{o}digo CAPES
001). Alexandre Dodonov agradece o apoio parcial do Conselho Nacional de
Desenvolvimento Cient\'{\i}fico e Tecnol\'{o}gico e Funda\c{c}\~{a}o de
Apoio \`{a} Pesquisa do Distrito Federal (FAPDF, projeto
00193-00001817/2023-43).

\section*{Disponibilidade de Dados}

Todo o conjunto de dados que dá suporte aos resultados deste 
estudo está disponível mediante solicitação ao autor correspondente 
[A. Dodonov].

\end{document}